\documentclass[aps,prd,superscriptaddress,twocolumn,10pt]{revtex4-1}

\usepackage{amssymb,amsmath,amsthm}
\usepackage{graphicx}
\usepackage{color}
\usepackage{url}
\usepackage{hyperref}
\hypersetup{colorlinks=false,pdfborder={0 0 0}}

\newcommand\EeV{\rm{EeV}}
\newcommand\ylm{Y_{\ell m}}

\newcommand{\uv}[1]{\hat{\mathbf{#1}}}

\newcommand{\e}[1]{\times10^{#1}}

\newcommand\veps\varepsilon
\newcommand\vphi\varphi

\newcommand\then\Rightarrow

\begin{document}

\title{Sensitivity of full-sky experiments to large scale cosmic ray anisotropies}
\author{Peter B. Denton}
\email{peterbd1@gmail.com}
\affiliation{Department of Physics \& Astronomy, Vanderbilt University, Nashville, TN, 37235, USA}
\author{Thomas J. Weiler}
\email{tom.weiler@vanderbilt.edu}
\affiliation{Department of Physics \& Astronomy, Vanderbilt University, Nashville, TN, 37235, USA}

\begin{abstract}
The two main advantages of space-based observation of extreme energy ($\gtrsim5\times10^{19}$~eV) cosmic rays (EECRs) over ground based
observatories are the increased field of view and the full-sky coverage with nearly uniform systematics across the entire sky.
The former guarantees increased statistics, whereas the latter enables a clean partitioning of the sky into spherical harmonics. 
The discovery of anisotropies would help to identify the long sought origin of EECRs.
We begin an investigation of the reach of a full-sky space-based experiment such as EUSO to detect anisotropies in the extreme-energy
cosmic-ray sky compared to ground based partial-sky experiments such as the Pierre Auger Observatory and Telescope Array.
The technique is explained here, and simulations for a Universe with just two nonzero multipoles, monopole plus either dipole or quadrupole, are presented.
These simulations quantify the advantages of space-based, all-sky coverage.
\end{abstract}

\date{\today}

\maketitle

\section{Introduction}
\label{sec:introduction}
\subsection{Previous Anisotropy Searches}
\label{ssec:previous}
No all-sky observatory has yet flown.
Consequently, the first full-sky large anisotropy search was based on combined northern and southern hemisphere 
ground-based data.
The respective data came from the SUGAR~\cite{0305-4616-12-7-015} and AGASA~\cite{Ohoka:1996ww} experiments,
taken over a 10~yr period.
Nearly uniform exposure to the entire sky resulted.  
No significant deviation from isotropy was seen by these experiments, even at energies beyond $4\e{19}$~eV~\cite{Anchordoqui:2003bx}. 
More recently, the Pierre Auger Collaboration carried out various searches for large scale anisotropies in the distribution of arrival
directions of cosmic rays above $10^{18}$ eV (an EeV)~\cite{Abreu:2011ve,Auger:2012an}.
At the energies exceeding $6\e{19}$~eV, early hints for a dipole anisotropy existed, but these hints have grown increasingly weaker 
in statistical strength~\cite{Anchordoqui:2011ks}.
The latest Auger study was performed as a function of both declination and right ascension (RA) in several energy ranges 
above $10^{18}$ eV.
Their results were reported in terms of dipole and quadrupole amplitudes.
Assuming that any cosmic ray anisotropy is dominated by dipole and quadrupole moments in this energy range, the Pierre Auger
Collaboration derived upper limits on their amplitudes.
Such upper limits challenge an origin of cosmic rays above $10^{18}$ eV from long lived galactic sources densely distributed in the
galactic disk~\cite{Abreu:2012ybu}.
In the $E>8$ EeV bin, they did report a dipolar signal with a $p$-value of $6.4\e{-5}$ 
(not including a ``look elsewhere" penalty factor)~\cite{ThePierreAuger:2014nja}.
Their cutoff of $\sim8$ EeV is above the galactic to extragalactic transition energy of $\sim1$ EeV, but still below the GZK cutoff energy
of $\sim55$ EeV.
Also, Telescope Array (TA), the largest cosmic ray experiment in the northern hemisphere, 
has reported a weak anisotropy signal above their highest energy cut of 57~EeV~\cite{Abbasi:2014lda}.

\subsection{Extreme Universe Space-Based Observatory}
\label{subsec:EUSO}
Proposals currently exist for all-sky, space-based cosmic-ray detectors such as the Extreme Universe Space Observatory (EUSO)~\cite{Adams:2013hqc} and
the Orbiting Wide-Angle Lens (OWL)~\cite{Krizmanic:2013pea}.
In addition, work is currently underway to combine datasets from two large ground based experiments, 
the Pierre Auger Observatory (Auger) in the 
southern hemisphere, and Telescope Array (TA) in the northern hemisphere~\cite{Deligny:2014fxa}.
This paper will use EUSO as the example for a full-sky observatory, but our conclusions will apply to any full-sky observatory.

EUSO is a down looking telescope optimized for near ultraviolet fluorescence produced by extended air showers in the atmosphere of the
Earth.
EUSO was originally proposed for the International Space Station (ISS), where it would collect up to 1000 cosmic ray (CR) events at and
above 55 \EeV\ ($1\,\EeV=10^{18}$~eV) over a $5$~year lifetime, far surpassing the reach of any ground based project.

It must be emphasized that because previous data were so sparse at energies which would be accessible to EUSO, upper limits on
anisotropy were necessarily restricted to energies below the threshold of EUSO.
EUSO expects many more events at $\sim 10^{20}$~eV, allowing an enhanced anisotropy reach.
In addition, EUSO would observe more events with a higher rigidity ${\cal R}=E/Z$, events less bent by magnetic fields; this may be
helpful in identifying point sources on the sky.

\subsection{Space-Based Advantages}
\label{subsec:advantages}
EUSO brings two new, major advantages to the search for the origins of extreme-energy (EE) CRs.
One advantage is the large field of view (FOV), attainable only with a space-based observatory.
With a $60^\circ$ opening angle for the telescope, the down pointing (``nadir'') FOV is
\begin{equation}
\pi(h_{\rm ISS}\tan(30^\circ))^2\approx h_{\rm ISS}^2\approx 150,000\,{\rm km}^2\,.
\label{eq:nadirFOV}
\end{equation}

We will compare the ability to detect large scale anisotropies at a space-based, full-sky experiment with that of a ground based
partial-sky experiment.
For reference, we will use the largest ground based cosmic ray observatory, the Pierre Auger Observatory~\cite{ThePierreAuger:2013eja}.
Auger has a FOV of 3,000 km$^2$.
Thus the proposed EUSO FOV, given in Eq.~\ref{eq:nadirFOV}, is 50 times larger for instantaneous measurements 
(e.g., for observing transient sources).
Multiplying the proposed EUSO FOV by an expected 18\% duty cycle, yields a time averaged nine-fold increase in acceptance for 
the EUSO design compared to Auger, at energies where the EUSO efficiency has peaked (at and above $\gtrsim 50-100$~\EeV). 
Tilting the telescope turns the circular FOV given in Eq.~\ref{eq:nadirFOV} into a larger elliptical FOV.
The price paid for ``tilt mode'' is an increase in the threshold energy of the experiment.

The second advantage of a space-based experiment over a ground based one is the coverage of the full-sky ($4\pi$~steradians) with
nearly constant exposure and consistent systematic errors on the energy and angle resolution, again attainable only with a space-based
observatory.
This paper compares full-sky studies of possible anisotropies to partial-sky studies.
The reach benefits from the $4\pi$~sky coverage, but also from the increased statistics resulting from the greater FOV.

In addition to the two advantages of space-based observation just listed, a third feature provided by a space-based mission may turn
out to be significant.  
It is the increased acceptance for Earth skimming neutrinos when the skimming chord transits ocean rather than land.
On this latter topic, just one study has been published~\cite{PalomaresRuiz:2005xw}.
The study concludes that an order of magnitude larger acceptance results for Earth skimming events transiting ocean compared to
transiting land.
Most ground based observatories will not realize this benefit, since they cannot view ocean chords, although those surrounded by ice
such as IceTop~\cite{Aartsen:2013lla} may also benefit from the ice as well.

The outline of this paper is the following:
We present the difference between partial-sky exposure and that of full-sky exposure in section~\ref{sec:comparison}.
In section~\ref{sec:tools} we review spherical harmonics as applied to a full-sky search for anisotropies, the power spectrum, and
anisotropy measures.
In section~\ref{sec:moments} we explain the particular interest in the dipole ($\ell=1$) and quadrupole ($\ell=2$) regions of
the spherical harmonic space as well as the techniques used here and in the literature for reconstructing the first two spherical
harmonics.
We also discuss the difficulties in differentiating dipoles and quadrupoles with partial-sky coverage; these difficulties are not present 
in full-sky coverage.
In section~\ref{sec:results} we present the results of our analysis for a pure dipole or quadrupole. Finally, some conclusions are
presented in section~\ref{sec:conclusion}.

\section{Comparison of Full-Sky Proposed EUSO to Partial-Sky Auger}
\label{sec:comparison}
The Pierre Auger Observatory is an excellent, largest-in-its-class, ground-based experiment.
However, in the natural progression of science, it is expected that eventually ground-based observation will be superseded by
space-based observatories.
EUSO is proposed to be the first-of-its-class, space-based observatory, building upon ground-based successes.

The two main advantages of a space-based observatory over a ground-based observatory 
are the greater FOV leading to a greater exposure at EE and the full-sky nature of the
orbiting, space-based observatory.
We briefly explore the advantage of the enhanced exposure first.
The 231--event sample published by Auger over 9.25 years of recording cosmic rays at and above $\sim52$~EeV~\cite{PierreAuger:2014yba}
allows us to estimate the flux at these energies.
The annual rate of such events at Auger is $\sim 231/9.25=25$.
For simplicity, we consider a 250 event sample for Auger, as might be collected over a full decade.

Including the suppressed efficiency of EUSO down to $\sim55$~\EeV\ reduces the factor of 9 relative to Auger down to a factor $\sim$6
for energies at and above 55\EeV. 
We arrive at the 450 event sample as the EUSO expectation at and above 55 \EeV\ after three years running in nadir mode (or, as is
under discussion, in tilt mode with an increased aperture but reduced PDM count).
A 750 event sample is then expected for five years of EUSO running in a combination of nadir and tilt mode.
Finally, the event rate at an energy measured by High Resolution Fly's Eye (HiRes) is known to significantly exceed that of Auger.
This leads to a five year event rate at EUSO of about 1000 events.
Thus, we consider the motivated data samples of 250, 450, 750, and 1000 events in the simulations that follow.

\begin{figure}
\centering
\includegraphics[width=\columnwidth]{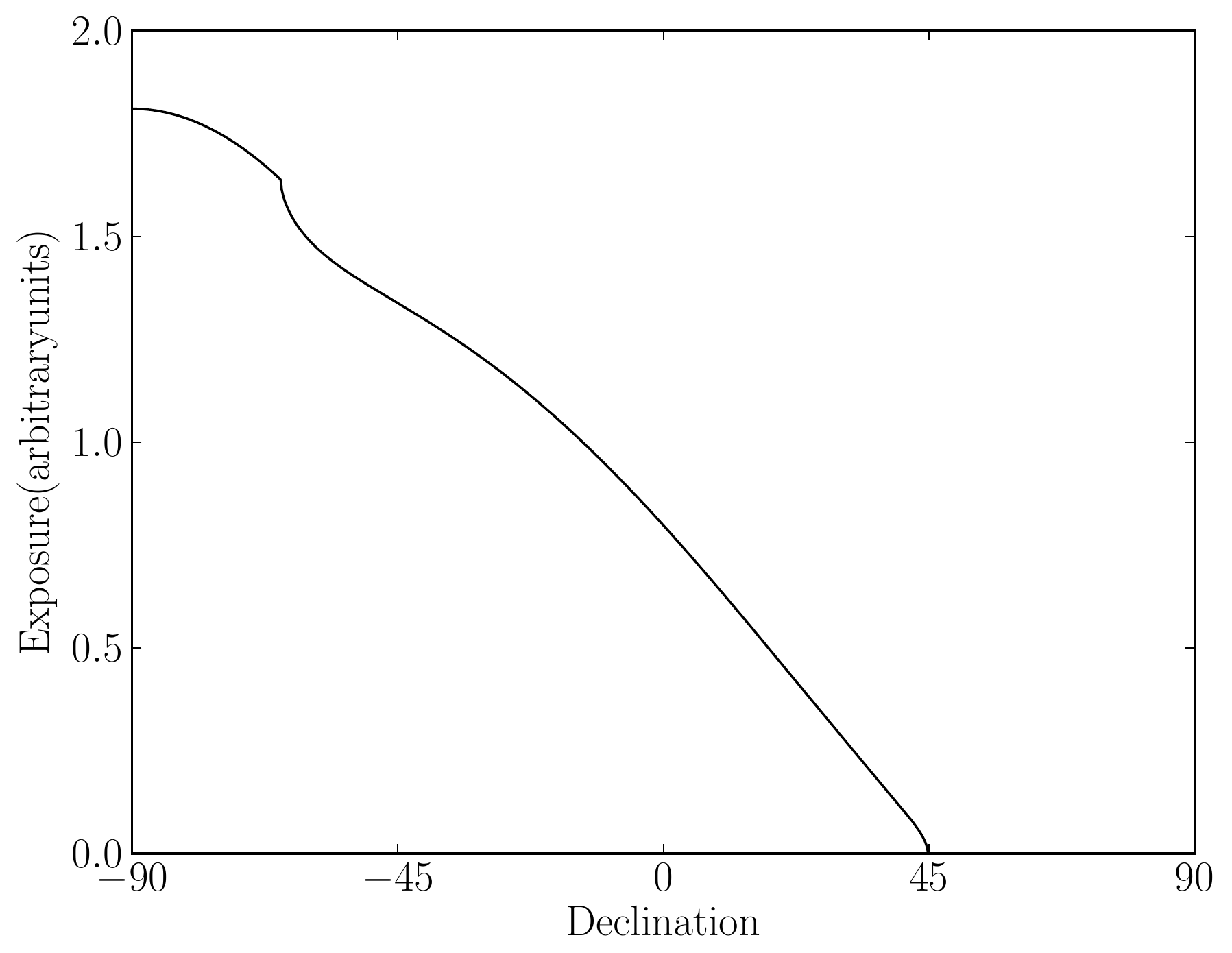}
\caption{Auger's exposure function normalized to $\int\omega(\Omega)d\Omega=4\pi$.
Note that the exposure is exactly zero for declinations $45^\circ$ and above.}
\label{fig:auger exposure}
\end{figure}

Now we turn to the $4\pi$~advantage.
Auger's exposure only covers part of the sky, and is highly nonuniform across even the part that it does see, as shown in
Fig.~\ref{fig:auger exposure}.
The relative exposure is given explicitly as~\cite{Sommers:2000us}
\begin{equation}
\begin{gathered}
\omega(\delta)\propto\cos a_0\cos\delta\sin\alpha_m+\alpha_m\sin a_0\sin\delta\\
\alpha_m=
\begin{cases}
0&{\rm for}\,\xi>1\\
\pi&{\rm for}\,\xi<-1\\
\cos^{-1}\xi\quad&{\rm else,}
\end{cases}\\
{\rm where\ } \xi\equiv\frac{\cos\theta_m-\sin a_0\sin\delta}{\cos a\cos\delta}\,,
\end{gathered}
\end{equation}
and where $\omega(\delta)$ is the relative exposure at declination $\delta$, $a_0=-35.2^\circ$ is Auger's latitude, $\theta_m=80^\circ$ is
the new~\cite{PierreAuger:2014yba} maximum zenith angle Auger accepts.
We have assumed that the detector is effectively uniform in
RA and that any variation (due to weather, down time, tilt of the machine, etc.) does not significantly affect 
the uniformity of the exposure in RA.
Auger recently modified their acceptance from $\theta_m=60^\circ\to80^\circ$ with the extension 
calculated using a different metric:
The $S(1000)$ technique is used for zenith angles $\theta\in[0^\circ,60^\circ]$, and the $N_{19}$ muon based technique 
is used for the new range, $\theta\in[60^\circ,80^\circ]$.
These inclined events extend Auger's reach up to a declination on the sky of $+45^\circ$,
as can be seen in Fig.~\ref{fig:auger exposure}.
In contrast, a space-based observatory such as EUSO would see in all directions with nearly uniform exposure.
Of course, Auger is an existing observatory, while EUSO is but a proposal.

\section{Tools for Anisotropy Searches}
\label{sec:tools}
\subsection{Spherical Harmonics on the Sky}
\label{ssec:spherical}
As emphasized by Sommers over a dozen years ago~\cite{Sommers:2000us}, a full-sky survey offers a rigorous expansion in spherical
harmonics,
of the normalized spatial event distribution $I(\Omega)$, where $\Omega$ denotes the solid angle parameterized by the pair of latitude
($\theta$) and longitude ($\phi$) angles,
\begin{equation}
I(\Omega)\equiv \frac{N(\Omega)}{\int d\Omega\, N(\Omega)} =\sum_{\ell=0}^\infty \, \sum_{|m| \le l} a_{\ell m}\,\ylm(\Omega)\,,
\label{eq:SphHarm}
\end{equation}
i.e., the set $\{\ylm\}$ is complete.
$N(\Omega)$ is the number of events seen in the solid angle $\Omega$.
The spherical harmonic coefficients, $a_{\ell m}$, then contain all the information about the distribution of events.
The set $\{\ylm\}$ is also orthonormal, obeying
\begin{equation}
\int d\Omega\;Y_{{\ell_1} {m_1}}(\Omega)\,Y_{{\ell_2}{m_2}}(\Omega)=\delta_{{\ell_1}{\ell_2}}\,\delta_{{m_1}{m_2}}\,.
\label{eq:ortho}
\end{equation}
We are interested in the real-valued, orthonormal $\ylm$'s, defined as
\begin{equation}
Y_{\ell m}(\theta,\phi)=
N(\ell,m)
\begin{cases}
P^\ell_m(x)(\sqrt{2}\cos(m\phi))&m>0\\
P_\ell(x)&m=0\\
P^\ell_m(x)(\sqrt{2}\sin(m\phi))\quad&m<0
\end{cases}\,,
\end{equation}
where, $P^\ell_m$ is the associated Legendre polynomial, $P_\ell=P^\ell_{m=0}$ is the regular Legendre polynomial, 
$x\equiv\cos\theta$, and the normalization-factor squared is $N^2(\ell,m) = \frac{(2\ell+1)(\ell-m)!}{4\pi\,(\ell+m)!}$.

The lowest multipole is the $\ell=0$ monopole, equal to the average full-sky flux and is fixed by normalization.
The higher multipoles ($\ell\ge 1$) and their amplitudes $a_{\ell m}$ correspond to anisotropies.
Guaranteed by the orthogonality of the $\ylm$'s, the higher multipoles when integrated over the whole sky equate to zero.

\begin{figure*}[t]
\centering
\newcommand\nodalwidth{0.197\textwidth}
\includegraphics[width=\nodalwidth]{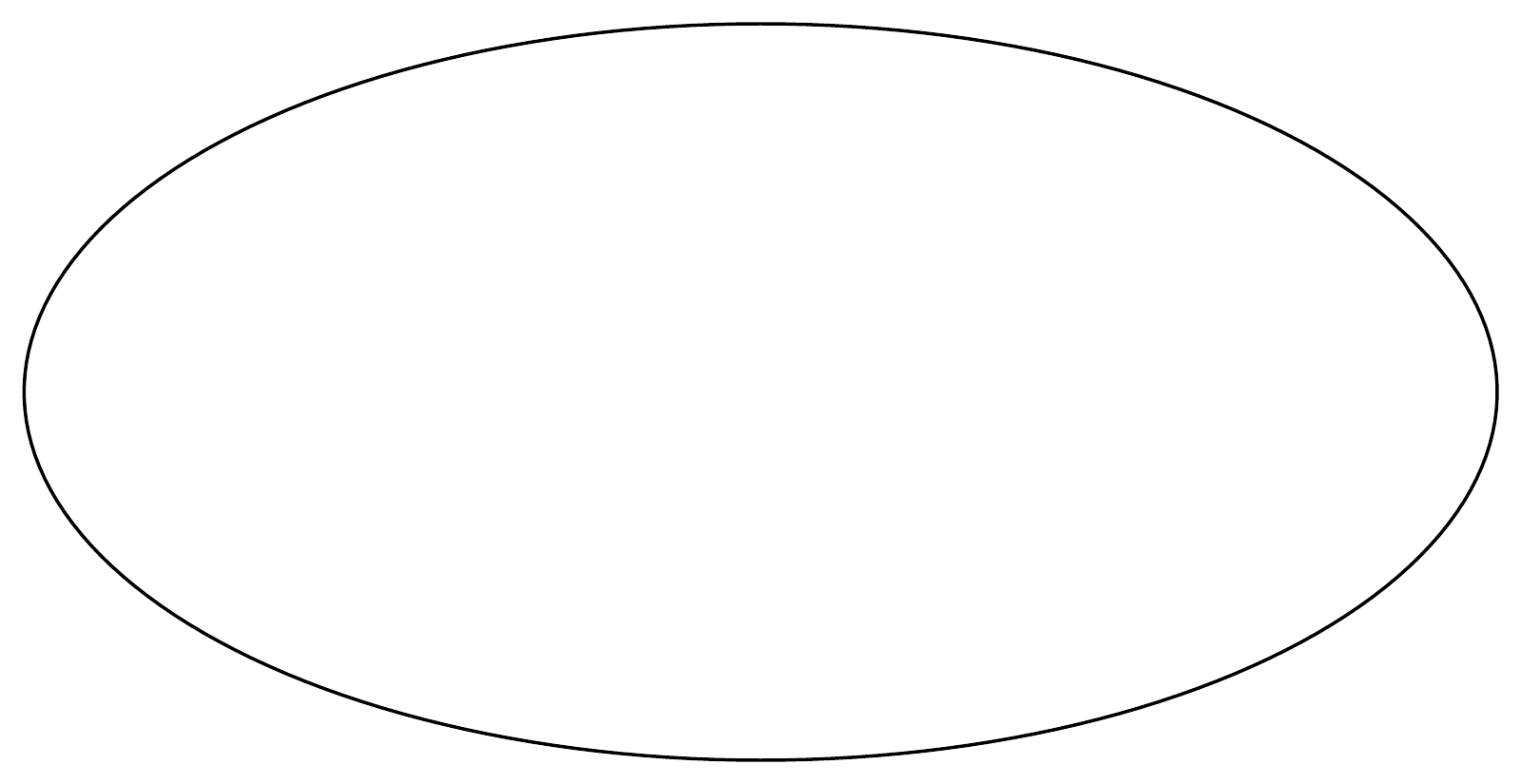}
\includegraphics[width=\nodalwidth]{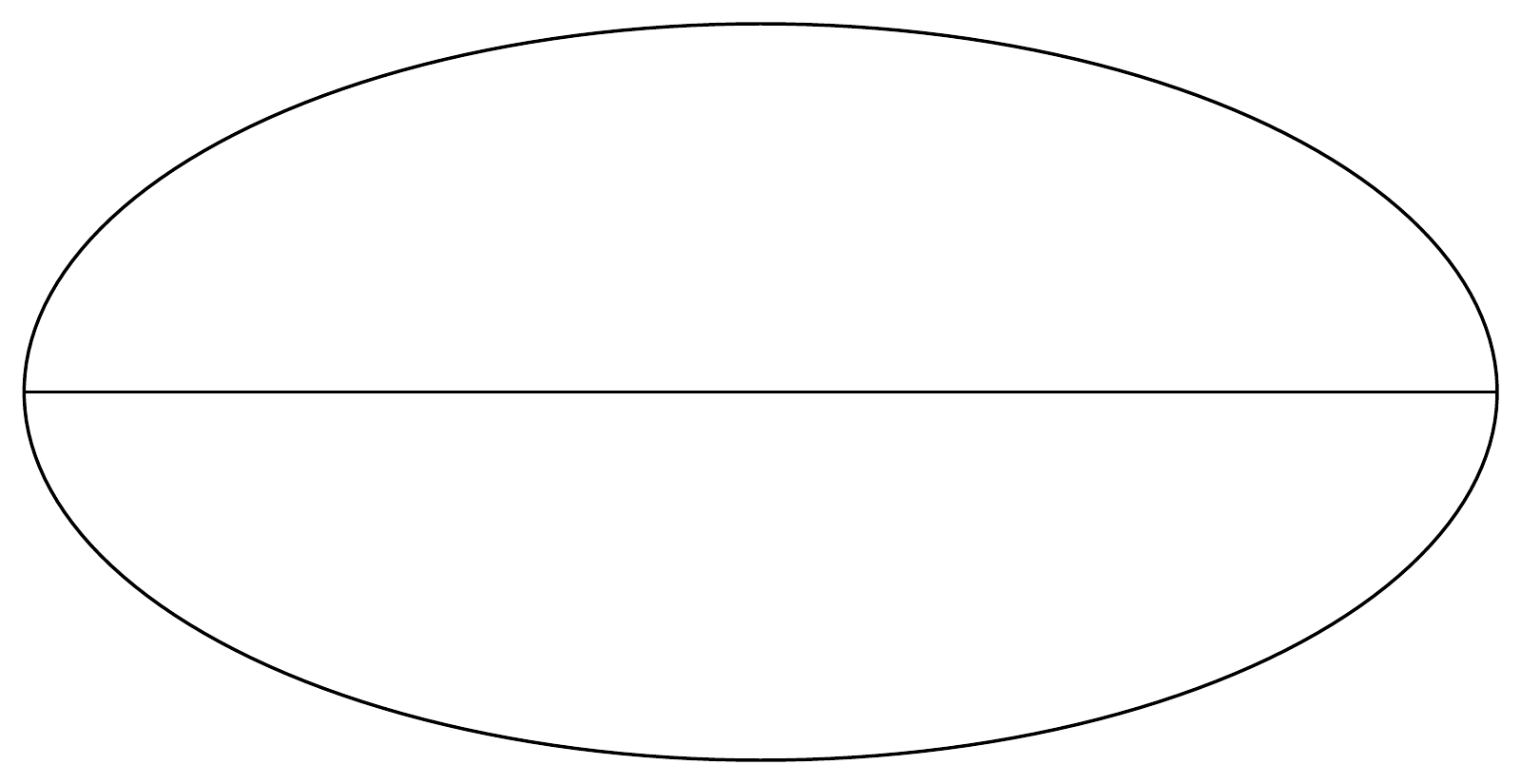}
\includegraphics[width=\nodalwidth]{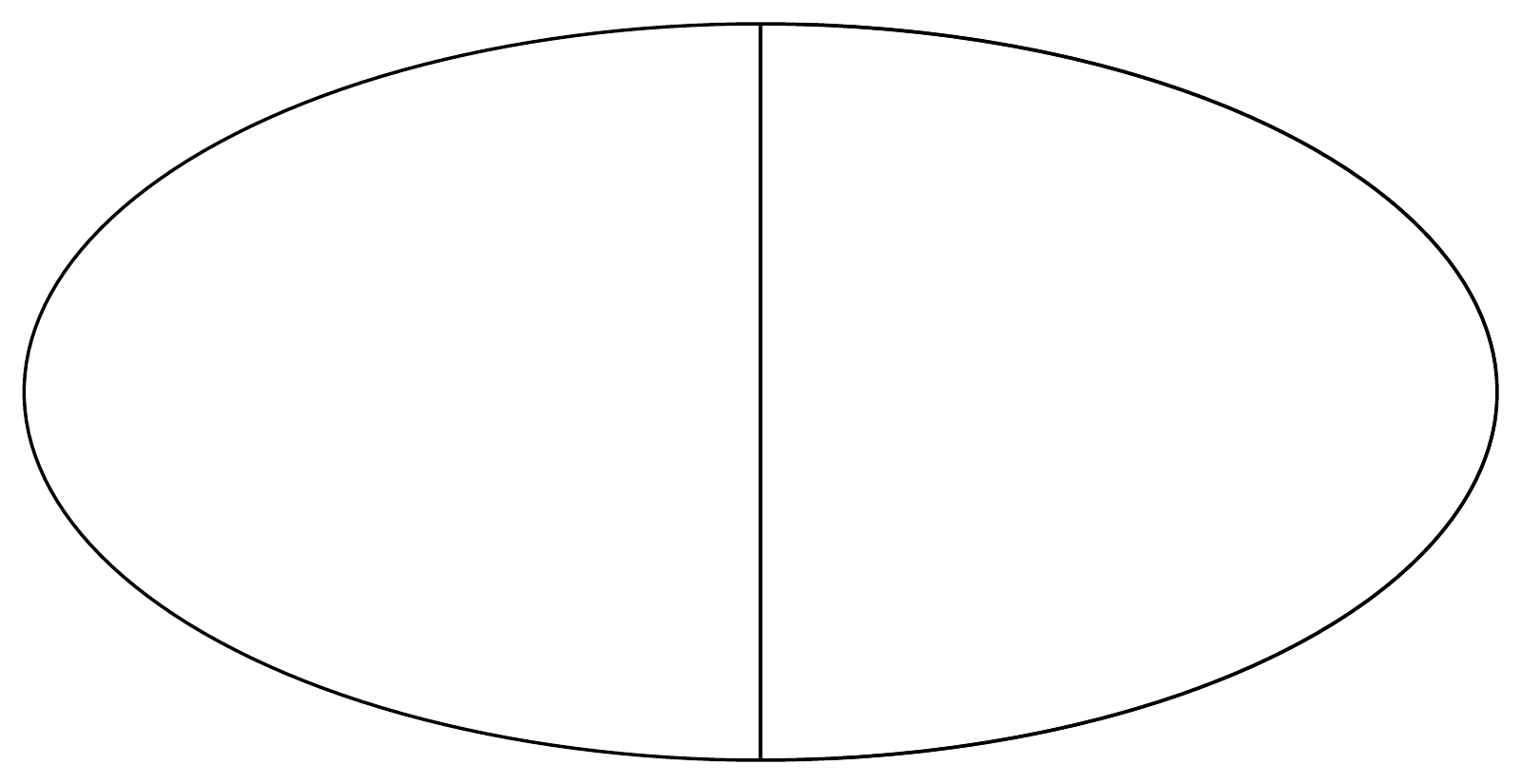}\\
\includegraphics[width=\nodalwidth]{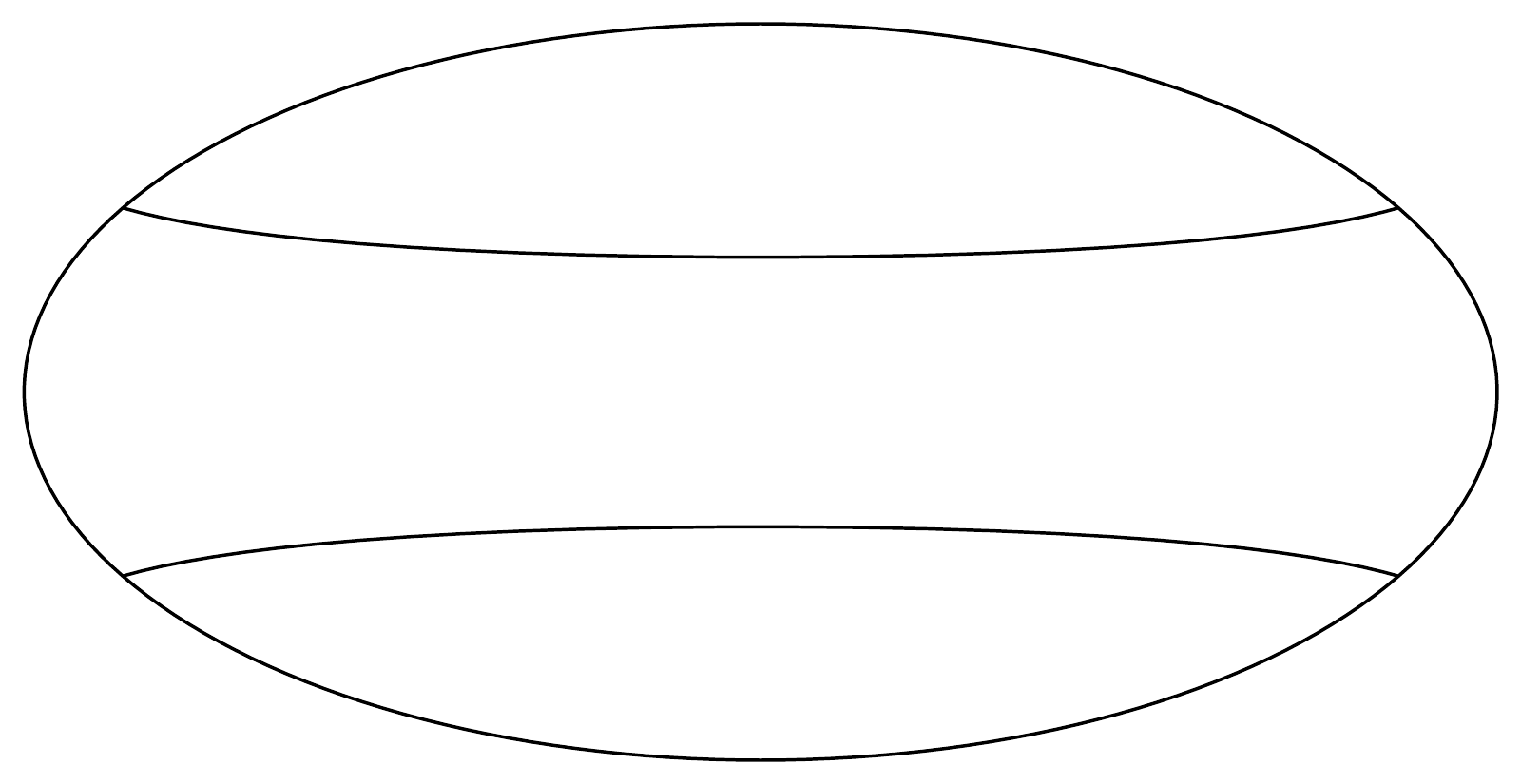}
\includegraphics[width=\nodalwidth]{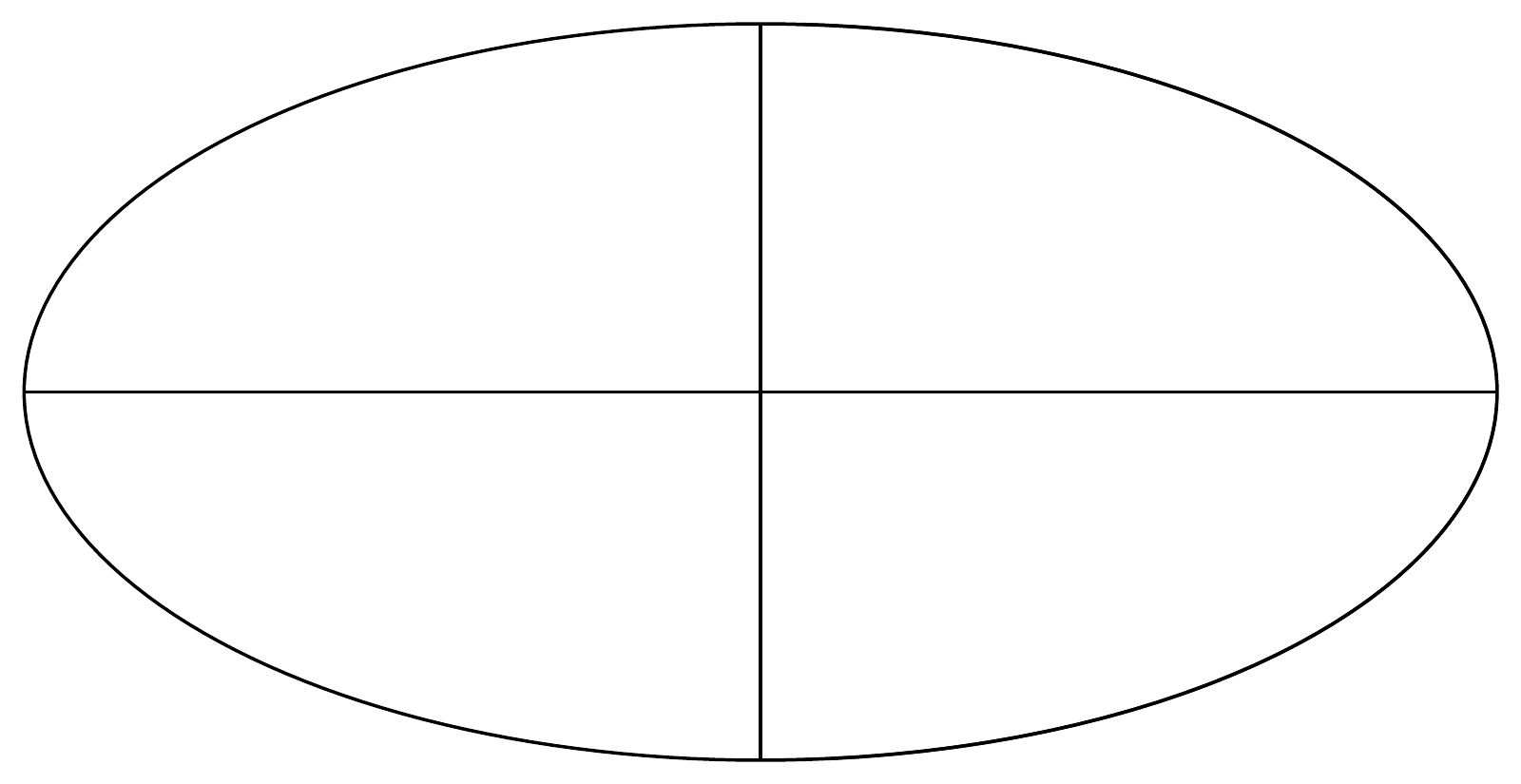}
\includegraphics[width=\nodalwidth]{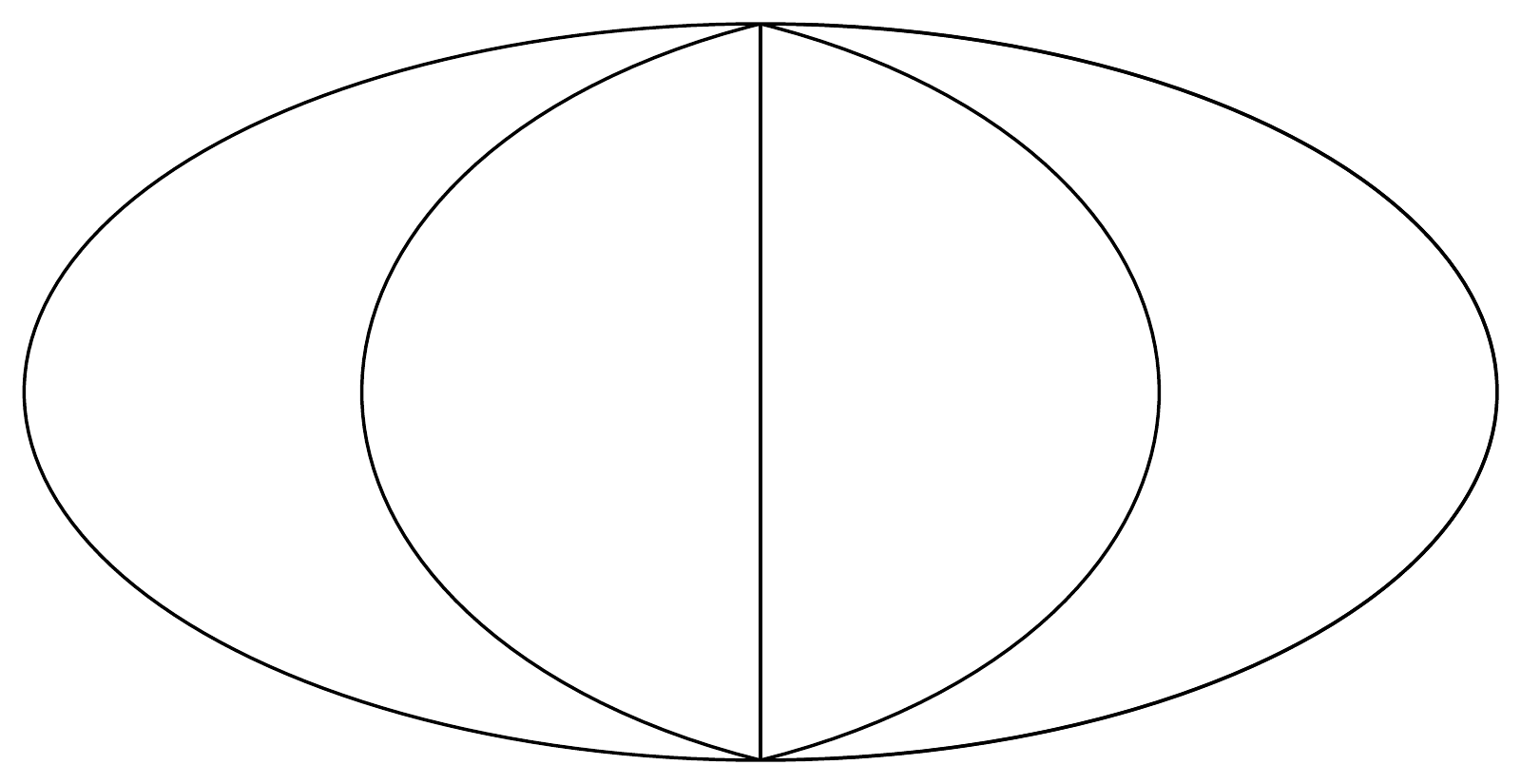}\\
\includegraphics[width=\nodalwidth]{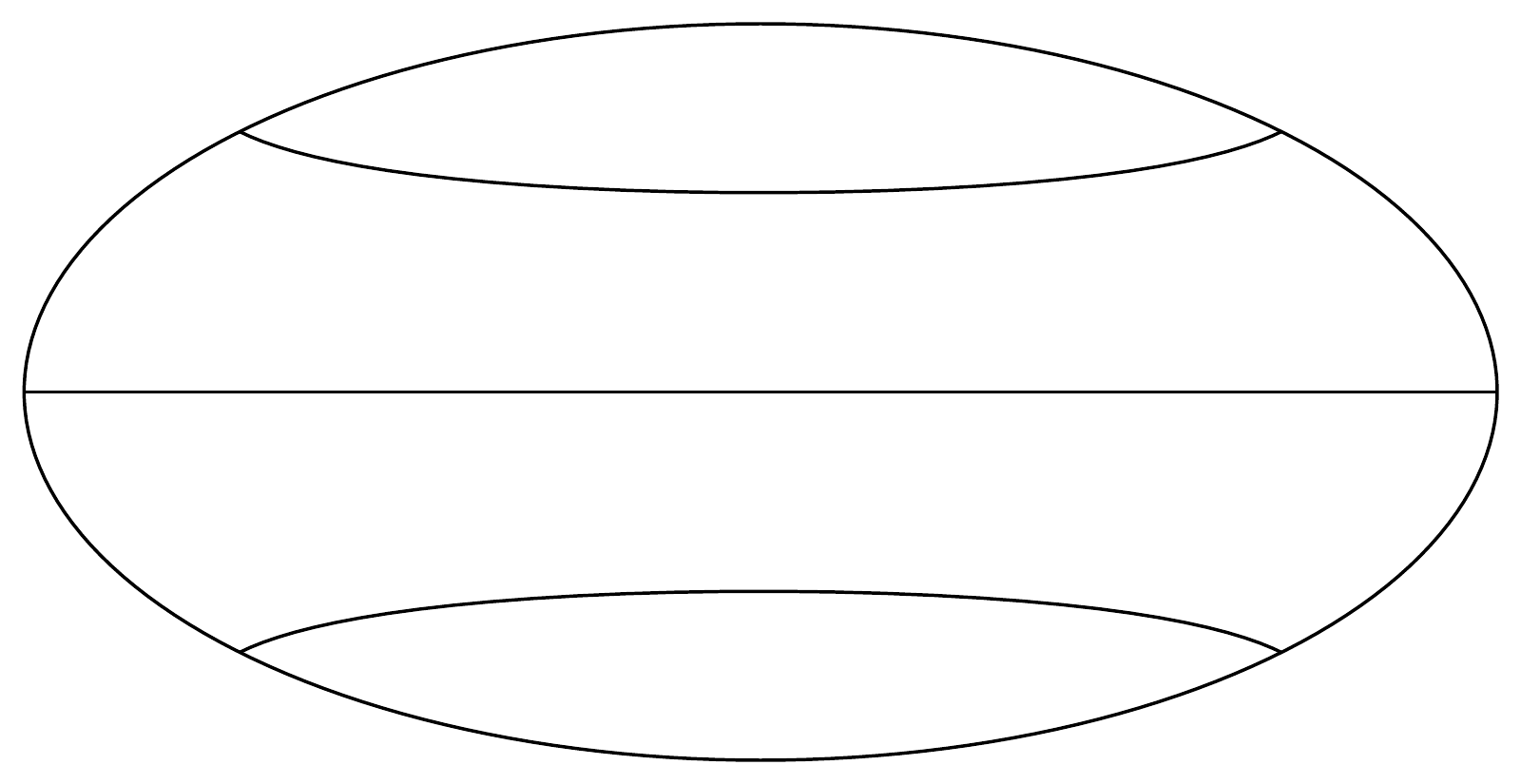}
\includegraphics[width=\nodalwidth]{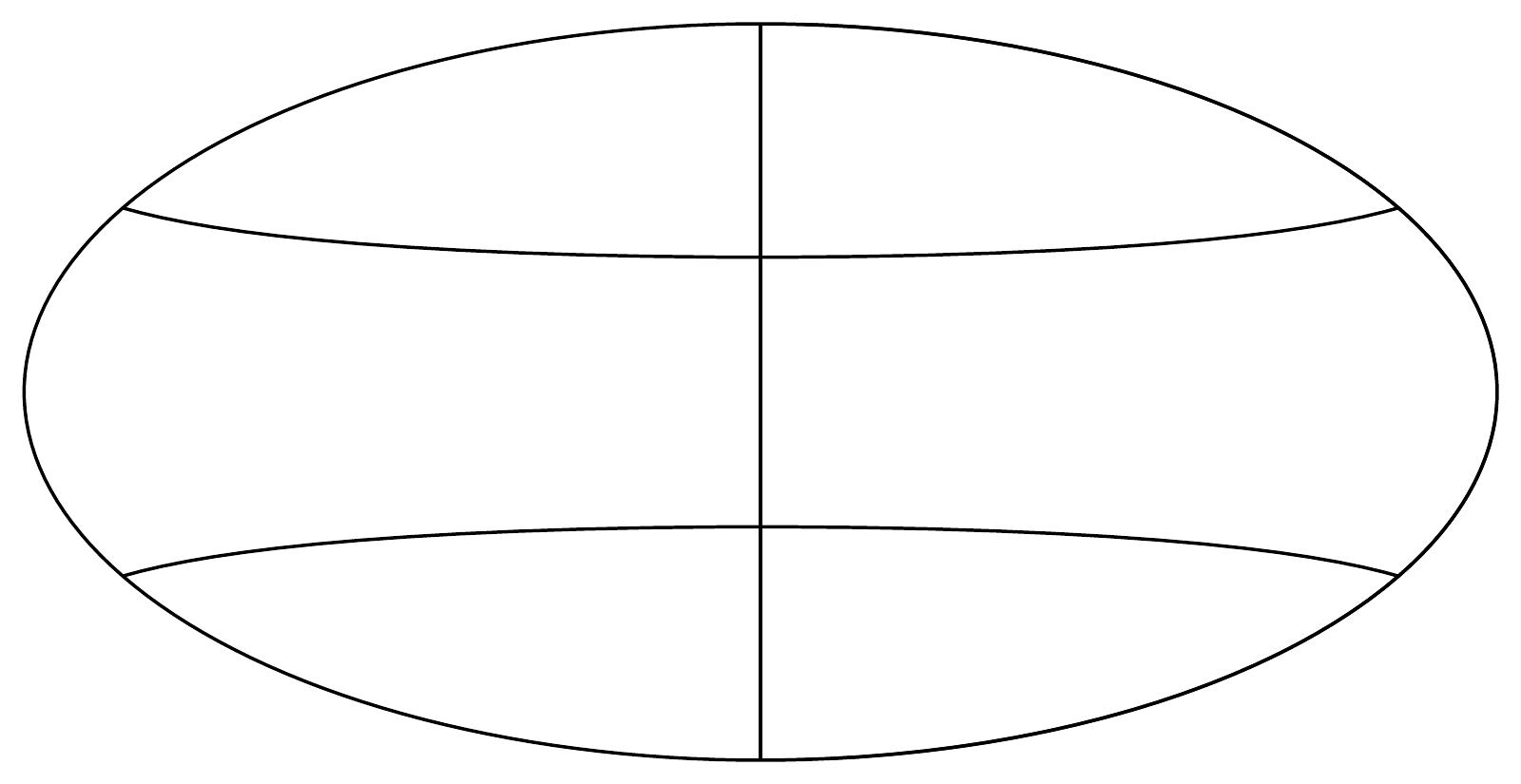} 
\includegraphics[width=\nodalwidth]{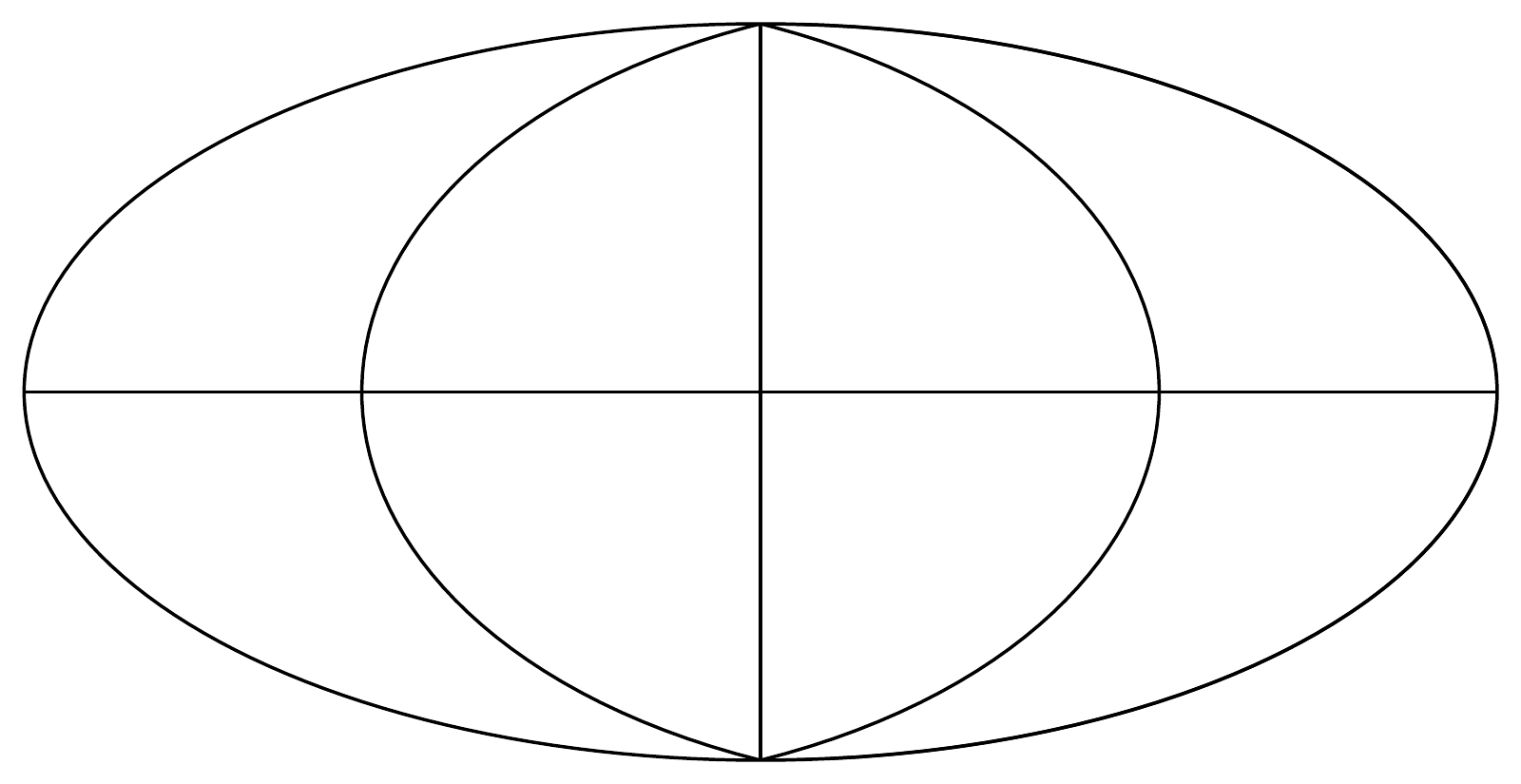}
\includegraphics[width=\nodalwidth]{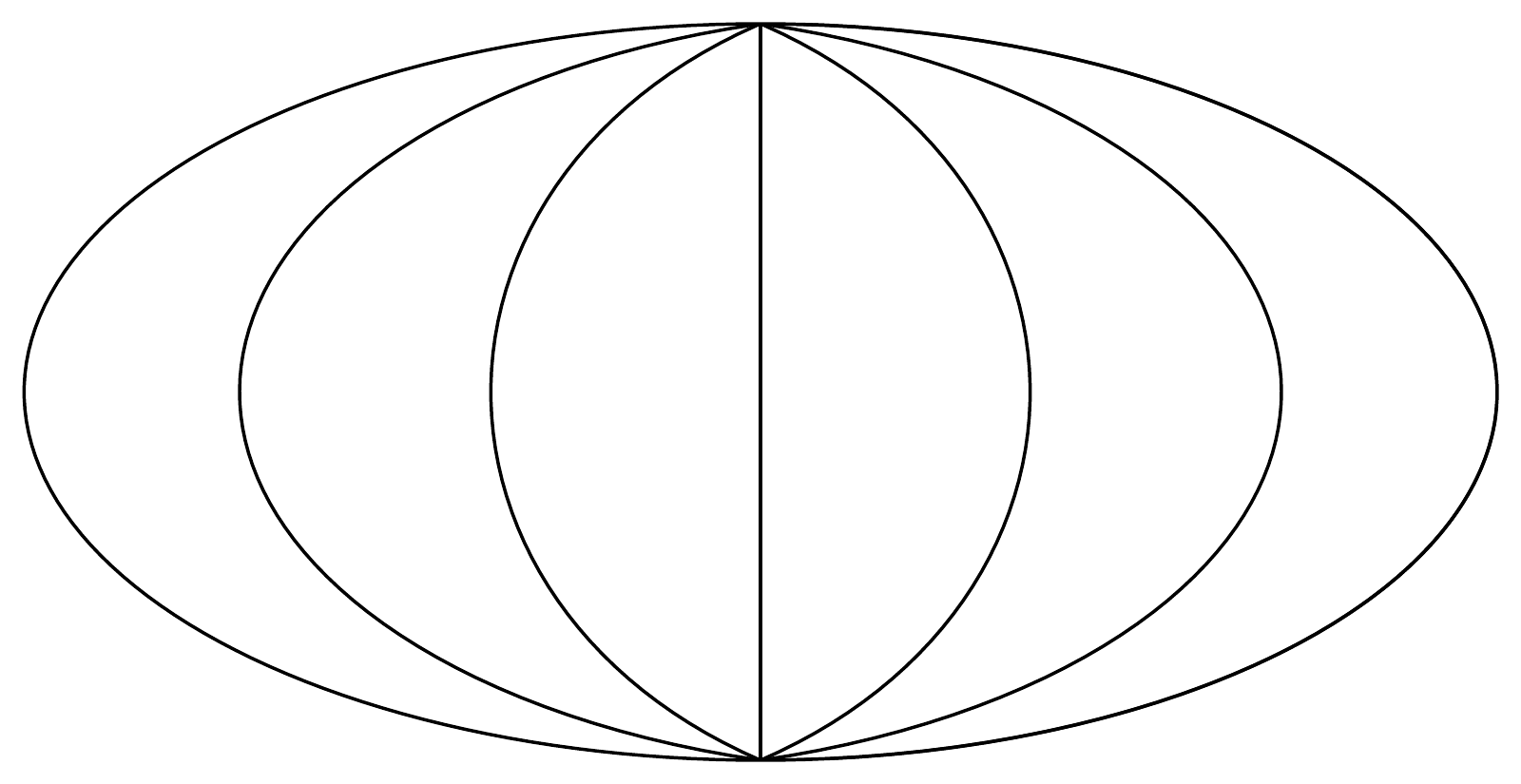}
\caption{Nodal lines separating excess and deficit regions of sky for various $(\ell, m)$ pairs.
The top row shows the $(0, 0)$ monopole, and the partition of the sky into two dipoles, $(1, 0)$ and $(1, 1)$.
The middle row shows the quadrupoles $(2, 0)$, $(2, 1)$, and $(2, 2)$.
The bottom row shows the $\ell=3$ partitions, $(3, 0)$, $(3, 1)$, $(3, 2)$, and $(3, 3)$.}
\label{fig:nodal}
\end{figure*}

A nonzero $m$ corresponds to $2\,|m|$ longitudinal ``slices'' ($|m|$ nodal meridians).
There are $\ell+1-|m|$ latitudinal ``zones'' ($\ell-|m|$ nodal latitudes).
In Fig.~\ref{fig:nodal}
we show the partitioning described by some low multipole moments.
Useful visualizations of spherical harmonics can also be found in Ref.~\cite{wiki:sphr.harm}.
The configurations with $(\ell,-|m|)$ are related to those with $(\ell,+|m|)$ by a longitudinal phase advance
$\phi\then\phi+\frac{\pi}{2}$, or $\cos\phi\then\sin\phi$.

\subsection{Power Spectrum}
\label{ssec:power spectrum}
The coefficients of the real-valued spherical harmonics, the $a_{\ell m}$'s, are real and frame-dependent.
To combat that problem, we use the power spectrum defined by
\begin{equation}
C_\ell\equiv\frac1{2\ell+1}\sum_{m=-\ell}^\ell a_{\ell m}^2\,.
\label{eq:power spectrum}
\end{equation}
That the $C_\ell$ should be rotationally (coordinate) invariant is not obvious.
Certainly the spherical harmonics coefficients, the $a_{\ell m}$'s given in Eq.~\ref{eq:SphHarm}, are coordinate dependent.
A simple rotation in the $\phi$ coordinate will change the $\sin\phi,\cos\phi$ part of the spherical harmonic for $m\neq0$ and a
rotation in the $\theta$ coordinate will change the associated Legendre polynomial part ($P_\ell^m(\theta)$) for $\ell\neq0$.
So only the $\ell=m=0$ monopole coefficient is coordinate independent. 
However, the power spectrum $C_\ell$ is invariant under rotations.
A recent derivation of this fact can be found in the appendices of Ref.~\cite{Denton:2014hfa}.

A simple approximation for the number of cosmic rays necessary to resolve power at a particular level is to count the number 
$N_Z(\ell,m)$ of nodal zones in each $\ylm$.
Each $\ylm$ has
\begin{equation}
N_Z(\ell,m)=
\begin{cases}
\ell+1&m=0\\
2|m|(\ell+1-|m|)\qquad&m\neq0
\end{cases}\,
\end{equation}
nodal zones.
The average over $m$ of the number nodal zones at a given $\ell$ is,
\begin{equation}
\langle N_Z(\ell)\rangle=\frac{\ell+1}{3(2\ell+1)}(2\ell^2+4\ell+3)\,.
\end{equation}
For low values of $\ell$, this returns the obvious results, $\langle N_Z(\ell=0)\rangle=1,\langle N_Z(\ell=1)\rangle=2$.
For large $\ell$, $\langle N_Z(\ell)\rangle\to\ell^2/3$.
If we make the simple assumption of requiring $\mathcal O(1)$ event per nodal zone to resolve a particular term in the power spectrum,
then, for large $\ell$ we require $\sim\ell^2/3$ to resolve $C_\ell$.
Thus, the rule of thumb is that our EUSO fiducial samples of 450, 750, and 1000 events can resolve the $C_\ell$'s up to 
an $\ell$-value of the mid-30's, mid-40's, and mid-50's, respectively, i.e., (using $\theta\sim \frac{90^\circ}{\ell}$) 
can resolve structures on the sky down to 2-$3^\circ$.  
A ground-based observatory, due to having fewer events and no full-sky coverage, would do much worse.
We note that the statistical error in angle estimated here for EUSO is well-matched to the expected systematic angular resolution
error~$\sim 1^\circ$ of EUSO.

\subsection{Anisotropy Measures}
\label{ssec:anisotropy measures}
Commonly, a major component of the anisotropy is defined via a max/min directional asymmetry,
\begin{equation}
\alpha\equiv\frac{I_{\max}-I_{\min}}{I_{\max}+I_{\min}}\in[0,1]\,.
\end{equation}
A dipole (plus monopole) distribution is defined by a dipole axis and an intensity map given by $I(\Omega)\propto1+A\cos\theta$, 
where $\theta$ is the angle between the direction of observation, denoted by $\Omega$, and the dipole axis.
This form contains a linear combination of the $Y_{1m}$'s.
In particular, a monopole term is required to keep the intensity map positive definite.
One readily finds that the anisotropy due to a dipole is simply $\alpha_D=A$.

A quadrupole distribution (with a monopole term but without a dipole term) is similarly defined, as
$I(\Omega)\propto1-B\cos^2\theta$.
In the frame where the $\uv z$ axis is aligned with the quadrupole axis, the quadrupole contribution is composed of just the $Y_{20}$
term.
In any other frame, this $Y_{20}$ is then related to all the $Y_{2m}$'s, by the constraint of rotational invariance of the $C_\ell$'s
mentioned above.
In any frame, one finds that the anisotropy measure is $\alpha_Q=\frac B{2-B}$,  and the inverse result is
$B=\frac{2\alpha}{1+\alpha}$.

\begin{figure*}
\centering
\includegraphics[width=0.48\textwidth]{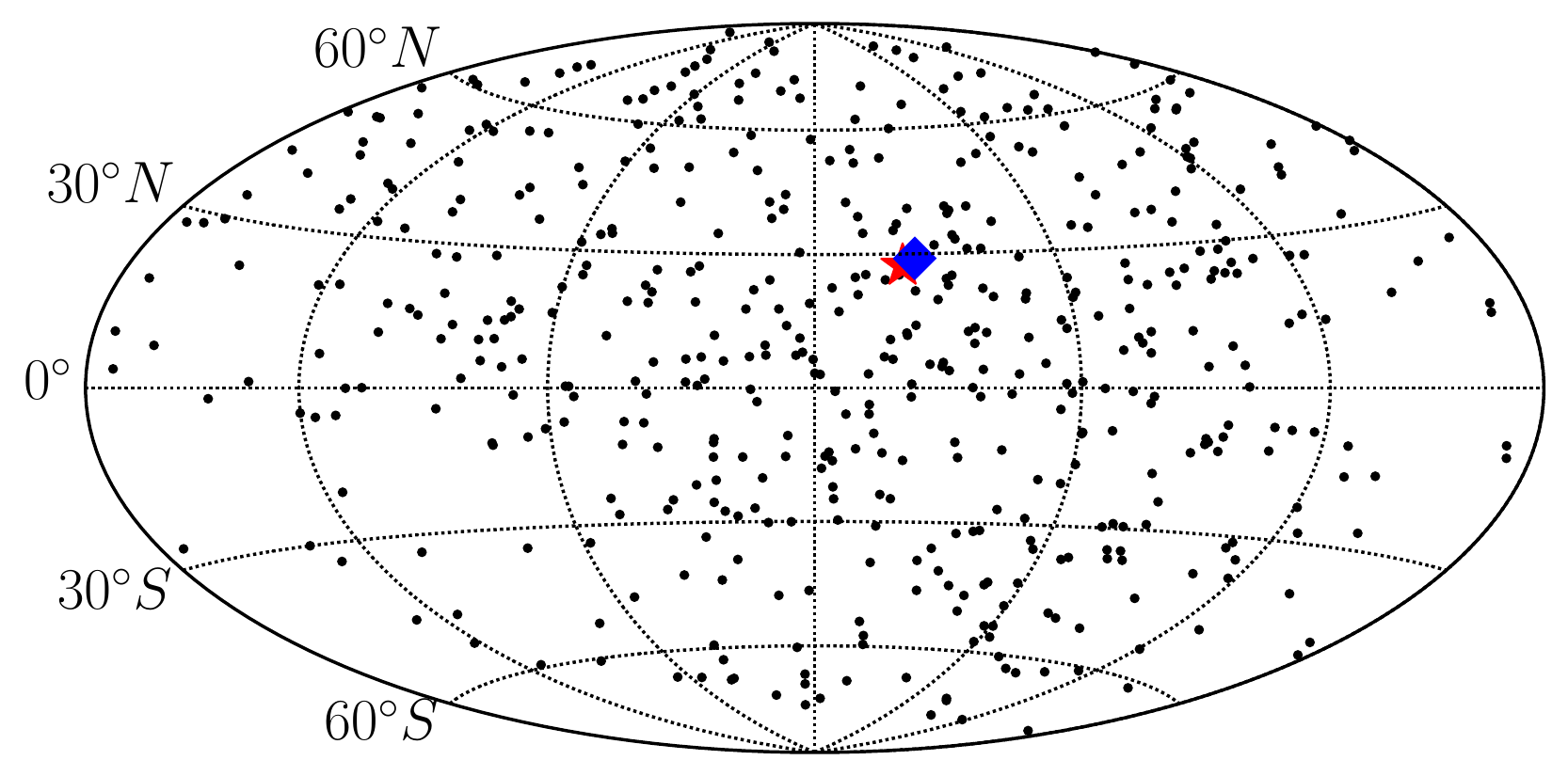}
\includegraphics[width=0.48\textwidth]{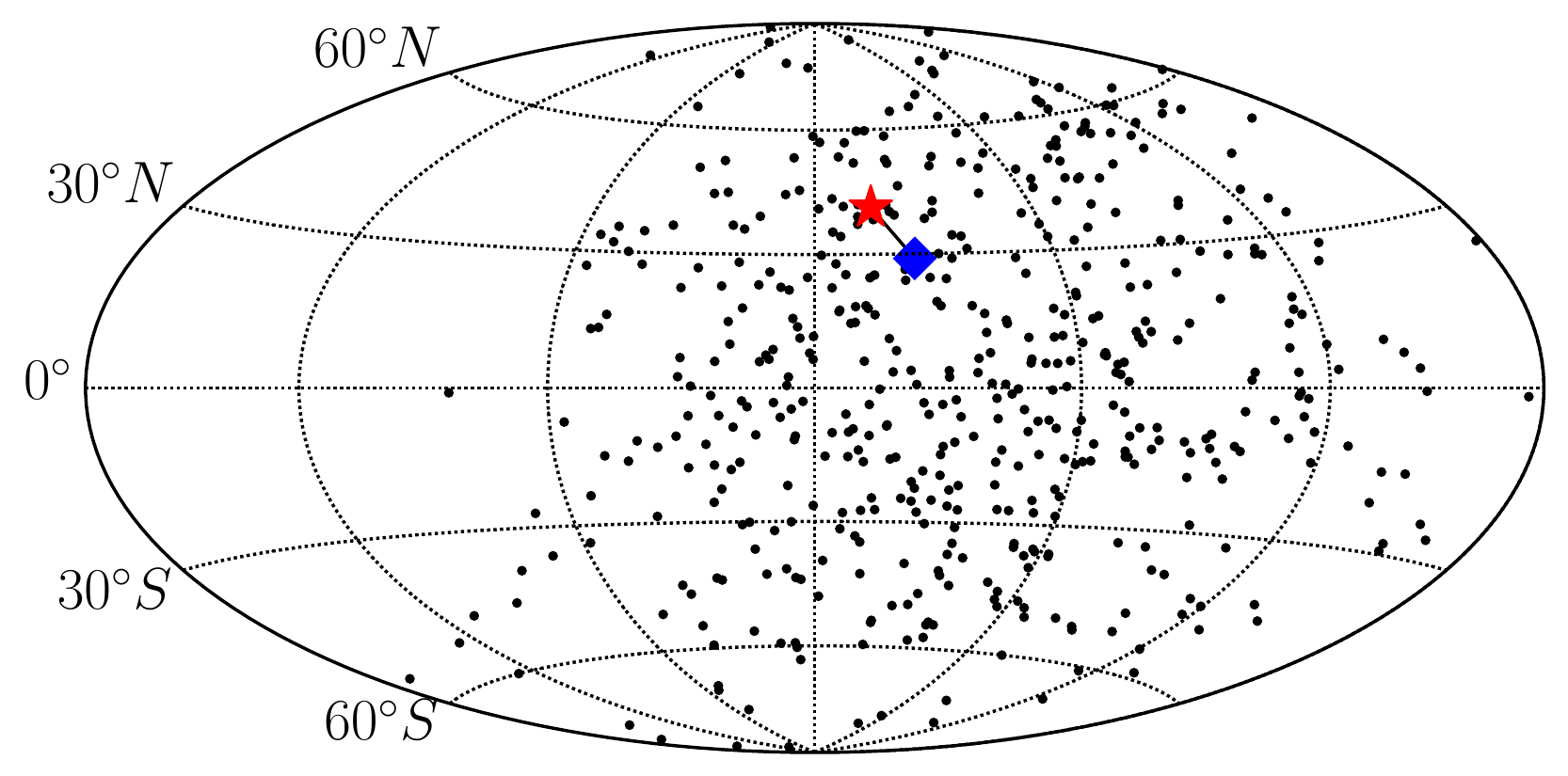}
\includegraphics[width=0.48\textwidth]{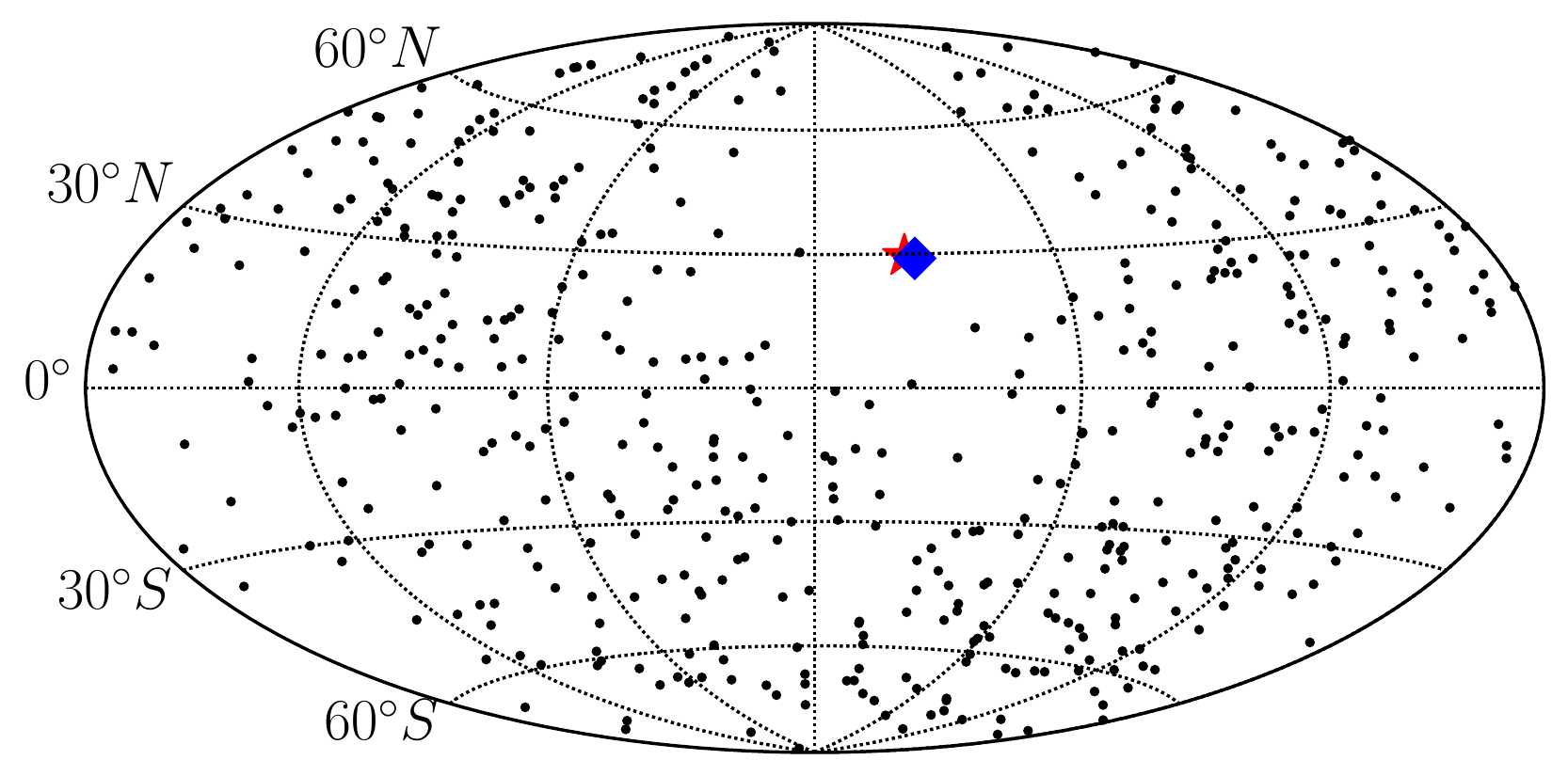}
\includegraphics[width=0.48\textwidth]{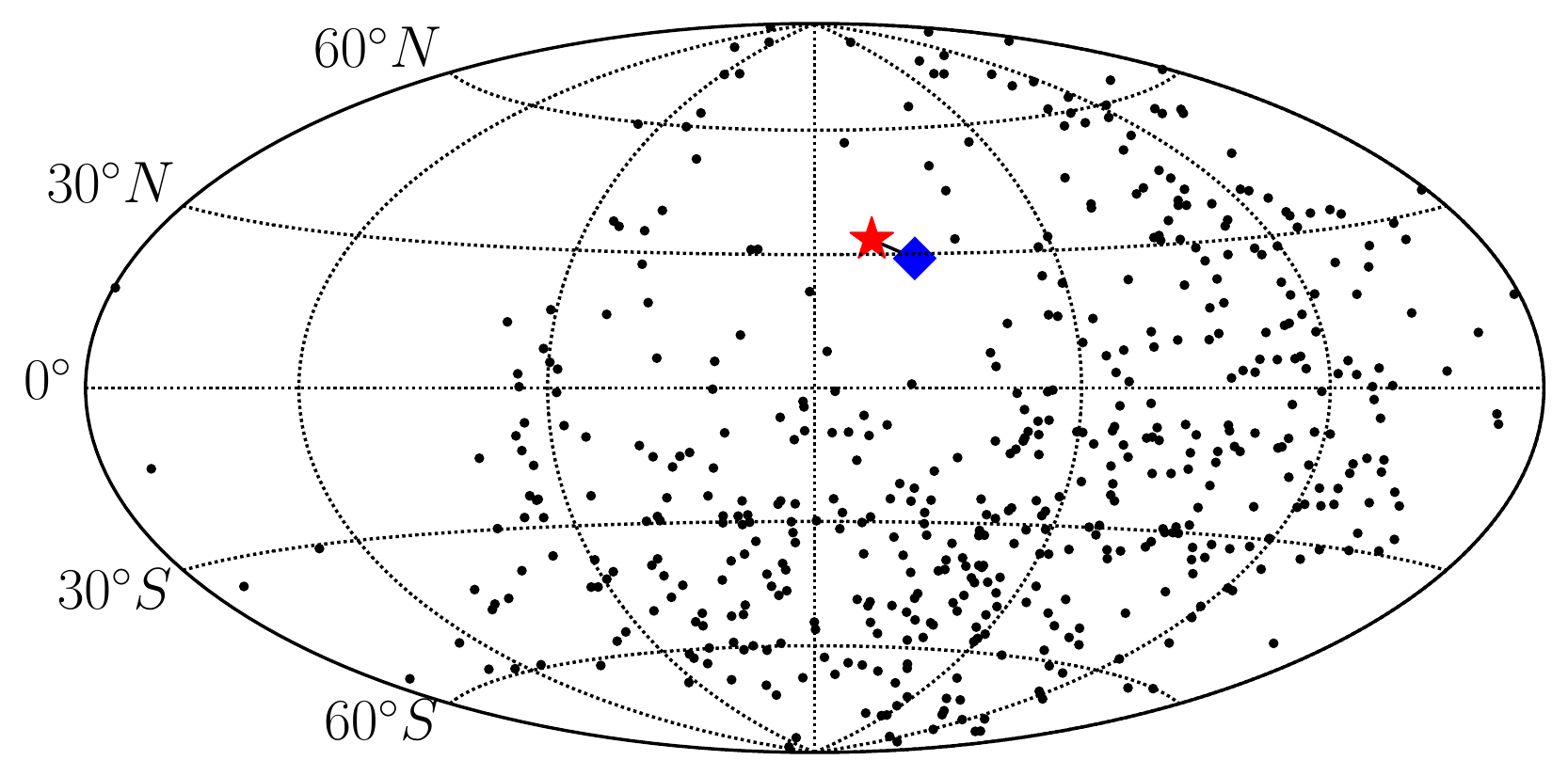}
\caption{Shown are sample sky maps of 500 cosmic rays.
The top row corresponds to the $\alpha_{D,{\rm true}}=1$ dipole,
while the bottom row corresponds to the $\alpha_{Q,{\rm true}}=1$ quadrupole distribution.
The left and right panels correspond to all-sky, space-based and partial-sky, ground-based coverage, respectively.
The injected dipole or quadrupole axis is shown as a blue diamond, and the reconstructed direction is shown as a red star.
We see that reconstruction of the multipole direction with an event number of 500 is 
excellent for an all-sky observatory (left panels) and quite good for partial-sky Auger (right panels).
In practice, $\alpha_D$ and $\alpha_Q$ are likely much less than unity, and the event rate for EUSO is expected to be $\sim 9$ times
that of Auger.
Both effects on the comparison of Auger and EUSO are shown in subsequent figures.}
\label{fig:skymaps}
\end{figure*}

Sample sky maps of dipole and quadrupole distributions are shown in Fig.~\ref{fig:skymaps} for both full-sky acceptance 
and for Auger's acceptance, along with the actual and reconstructed symmetry axes.

\section{Reconstructing Spatial Moments}
\label{sec:moments}
\subsection{Reconstructing a Dipole Moment}
\label{ssec:reconstructing dipole}
Dipoles excite the specific spherical harmonics corresponding to $Y_{1m}$, with the three $Y_{1m}$'s proportional to $\uv x$, $\uv y$,
and $\uv z$.
A dipolar distribution is theoretically motivated by a single distant point source producing the majority of EECRs 
whose trajectories are subsequently smeared by galactic and extragalactic magnetic fields.

With full-sky coverage it is straightforward to reconstruct the dipole moment so long as the exposure function 
is always nonzero (and possibly nonuniform).
For the full-sky case (EUSO), we use the description described in~\cite{Sommers:2000us}, 
which even allows for a nonuniform exposure,
provided that the exposure covers the full-sky.

\begin{figure*}
\centering
\includegraphics[width=0.48\textwidth]{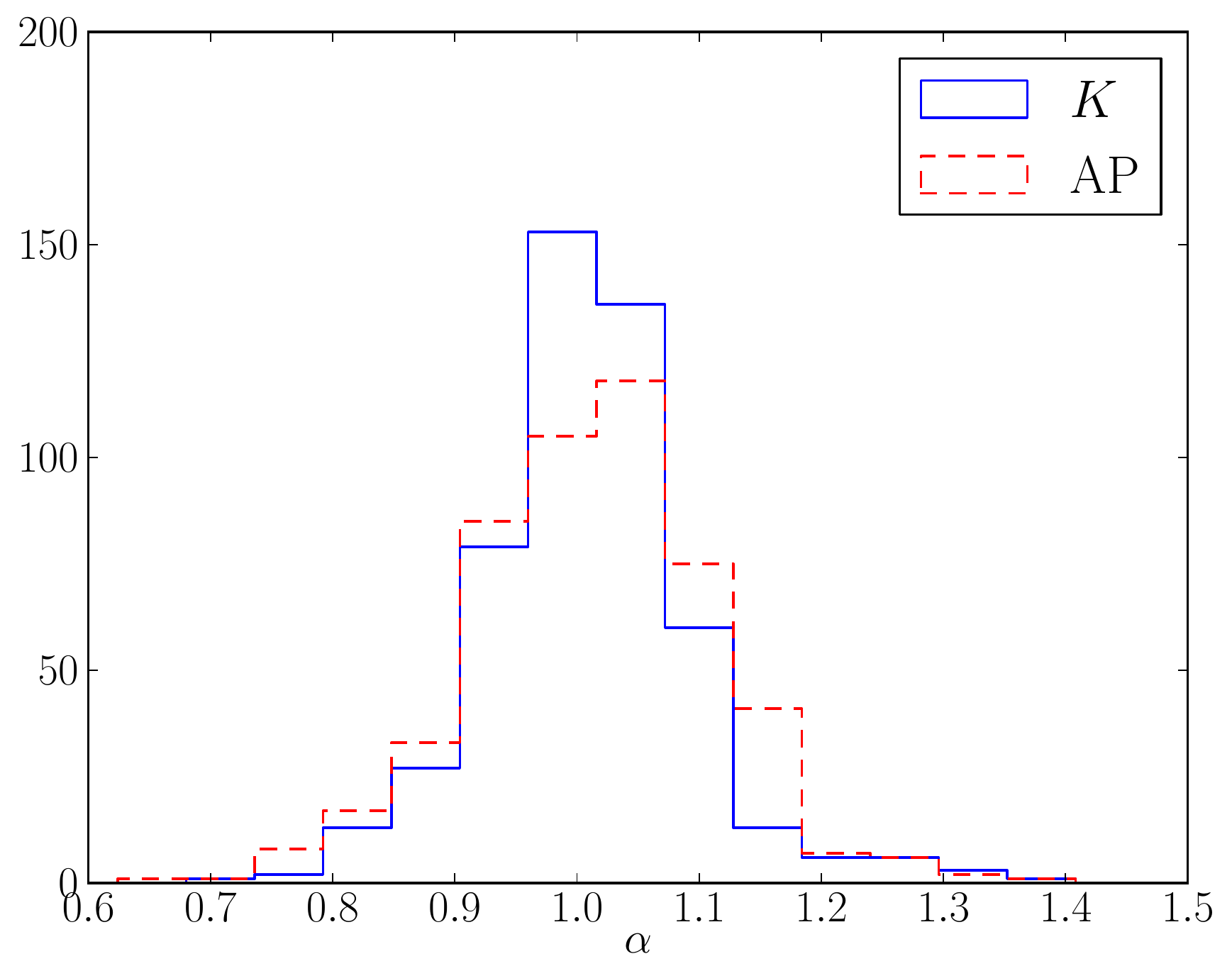}
\includegraphics[width=0.48\textwidth]{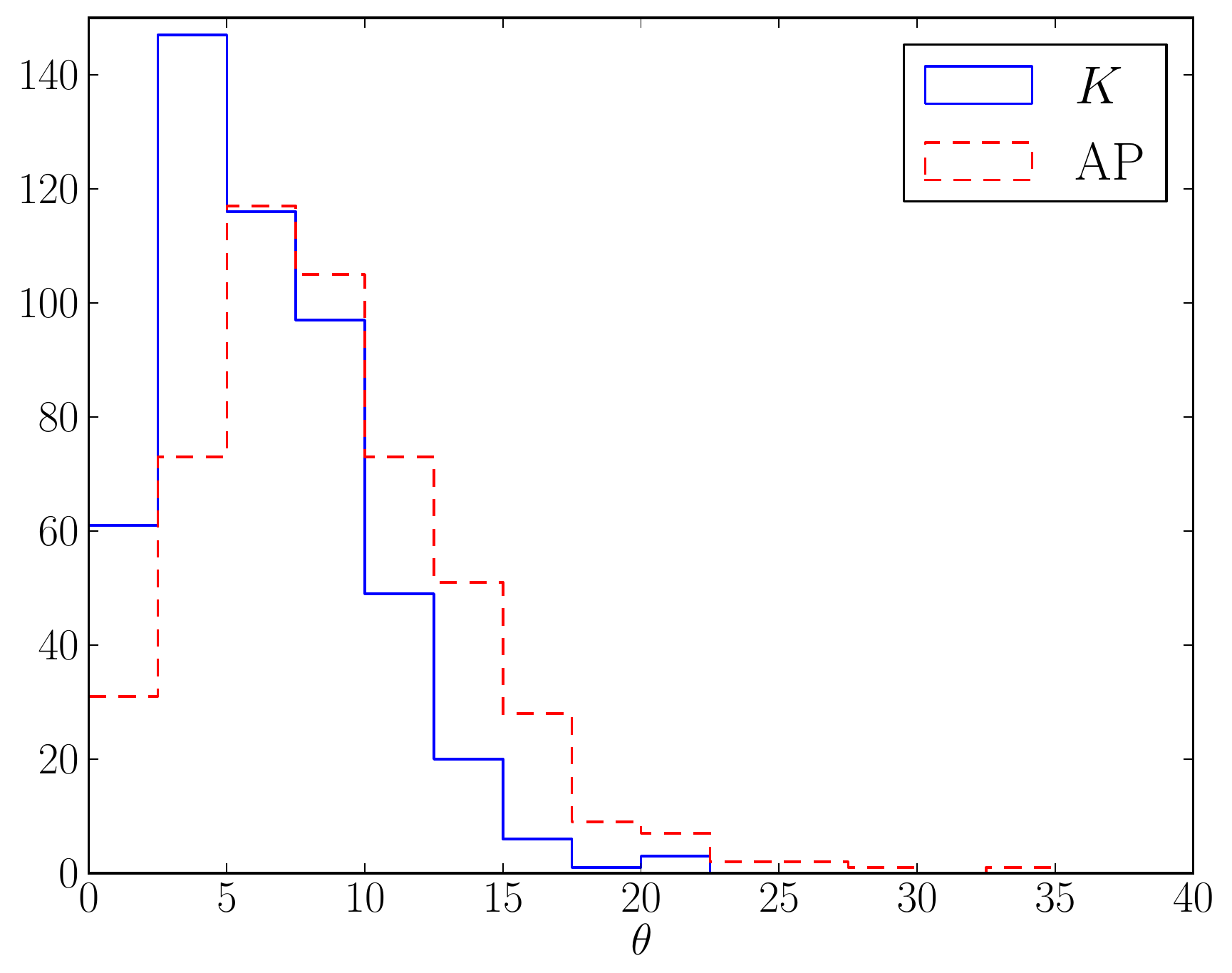}
\caption{We simulated 500 cosmic rays with a dipole of amplitude of $\alpha_{D,{\rm true}}=1$ pointing in a random direction, 
and with the Auger exposure.
Then we reconstructed both the direction and the dipole amplitude 500 times 
with both reconstruction techniques (AP and $K$-matrix).
In the left panel we show a histogram of the reconstructed values of $\alpha_{D,{\rm true}}$, 
and in the right panel the angle (in degrees) between
the correct dipole direction and the reconstructed direction.}
\label{fig:APvK}
\end{figure*}

Reconstructing any anisotropy, including the dipole, with partial-sky exposure is challenging.
One approach for dipole reconstruction is that presented in~\cite{Aublin:2005nv}. 
We refer to this approach as the AP method.
We note that this AP approach becomes very cumbersome for reconstructing the quadrupole and higher multipoles.
Another approach for reconstructing any $\ylm$ with partial coverage is presented in~\cite{Billoir:2007kb},
which we refer to as the $K$-matrix approach.
For the dipole with partial-sky coverage, we compare these two approaches to determine which optimally 
reconstructs a given dipole distribution.  Our result is seen in Fig.~\ref{fig:APvK}.
We consider 500 cosmic rays with a dipole distribution of magnitude $\alpha_{D,\rm{true}}=1$ 
oriented in a random direction.
Using Auger's exposure map, we then reconstruct the strength of the dipole using each method.
This entire process was repeated 500 times.
The reconstructed values of $\alpha_D$ are $\alpha_{D,\rm{rec}}=1.013\pm0.101$ and $\alpha_{D,\rm{rec}}=1.009\pm0.084$ for
the AP and $K$-matrix approaches, respectively, where the uncertainty is statistical.
The mean angles between the actual dipole direction and the reconstructed dipole direction are $\theta=8.82^\circ$ and
$\theta=6.41^\circ$ for the AP and $K$-matrix approaches respectively.
The the results of the two approaches are comparable. 
Since the $K$-matrix approach does slightly better than the AP method, 
we will use the $K$-matrix approach for partial-sky dipole reconstructions in what follows.

\subsection{Reconstructing a Quadrupole Moment}
\label{ssec:reconstructing quadrupole}
The physically motivated quadrupoles are the spherical harmonics corresponding to $Y_{20}\propto3z^2-1$.
$Y_{20}$ represents an anisotropy that is maximal along the equator and minimal along the poles 
(or, depending on the sign of $a_{20}$, the opposite).
Such a distribution is motivated by the presence of many sources distributed along a plane, such as is the case with the supergalactic
plane.
As a real, physical example of a well-known source-distribution distributed with a quadrupole contribution, we calculate the power
spectrum as might be seen at the Earth for the 2MRS catalog of the closest 5310 galaxies above a minimum intrinsic
brightness~\cite{Huchra:2011ii}.
The catalog contains redshift information and contains all galaxies above a minimum intrinsic brightness out to $z=0.028\sim120$ Mpc.
As such, it is reasonable to suppose that EECRs come from these galaxies and, for simplicity, we implement uniform flux from each
galaxy.

\begin{figure*}
\includegraphics[width=\columnwidth]{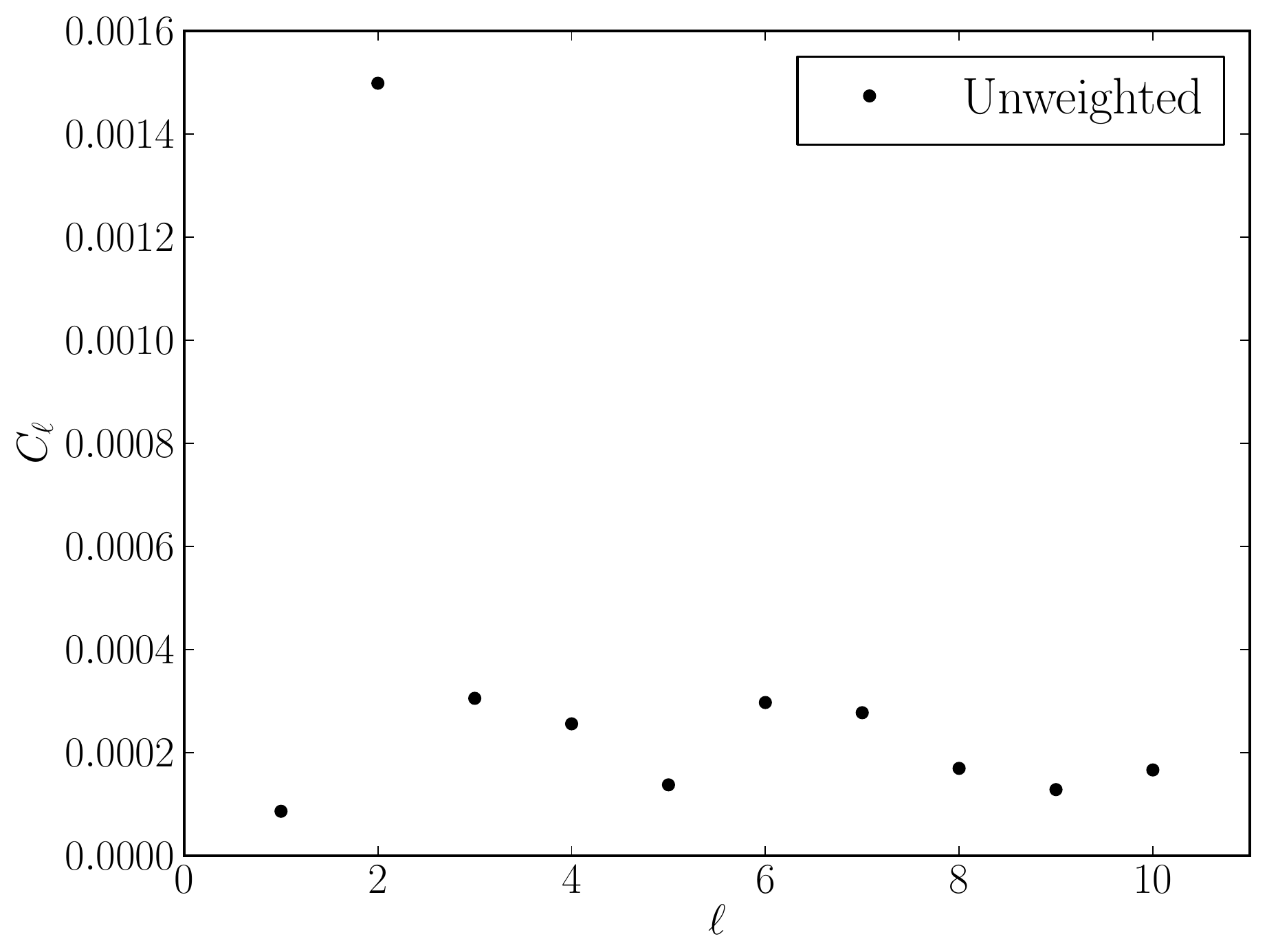}
\includegraphics[width=\columnwidth]{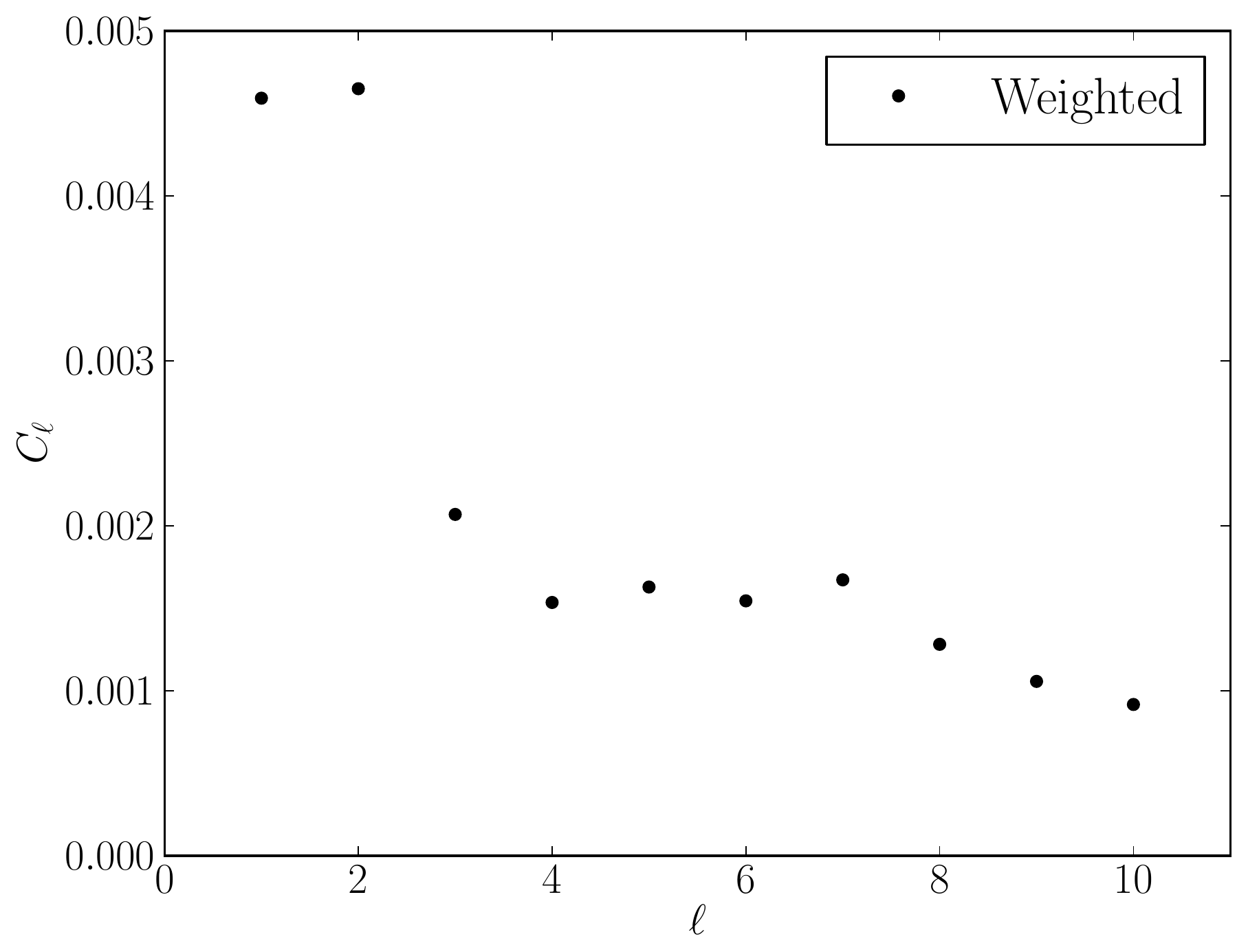}
\caption{The power spectrum (see Eq.~\ref{eq:power spectrum}) for nearby galaxies out to $z=0.028$ (to $d=120$ Mpc) based on their
positions (left panel), and weighted according to $1/d^2$ (right panel).
The 2MRS catalog~\cite{Huchra:2011ii} includes a cut on Milky Way latitudes $|b|<10^\circ$ which is accounted for in the calculation of
the power spectrum.
$C_2$ is large because the galaxies roughly form a planar (quadrupolar) structure; $C_1$ in the right panel is large because 
we are not in the center of the super cluster, thereby inducing a dipole contribution.
(The relative scale between the ordinates of the two figures carries no information.) }
\label{fig:Clgals}
\end{figure*}

In the left panel of Fig.~\ref{fig:Clgals} we show the power spectrum that results for the known physical locations of these galaxies.
In the right panel we show the power spectrum that results when each galaxy is weighted by the number of events expected from it, i.e.,
by the inverse-square of the distance to the galaxy, a $1/d^2$ weighting. 
We remark that for the closest $\sim200$~galaxies, the distance to each galaxy is known better from the direct ``cosmic distance
ladder'' approach than it is from the redshift, and we use these direct distance.
For the farther galaxies, direct distances are less reliable, and we use the redshift-inferred distances.
In this way, we also avoid any (possibly large) peculiar-velocity contributions to the redshifts of the nearer galaxies.

It is instructive to compare the two panels.
Without the $1/d^2$ weighting (left panel), the intrinsic quadrupole nature of the distribution of 2MRS galaxies dominates the power
spectrum; $C_2$ exceeds the other $C_\ell$'s in the panel by a factor of $\gtrsim5$.
In the right panel, galaxies are weighted by their apparent fluxes so the closest galaxies dominate.
The large dipole is due to the proximity of Cen A, and the fact that the next closest galaxy, M87, is $\sim4$ times farther from the
Earth.
When determinations of the $C_\ell$'s are made, it is likely to be the dipole and quadrupole that will first emerge from the data
based on the distributions of nearby galaxies.
This quantifiably motivates our choice made in this paper to examine the dipole and quadrupole anisotropies.
While the actual distribution is likely a combination of dipole and quadrupole components, throughout this paper we consider the
simpler cases where the distribution of sources has either a pure dipole anisotropy or a pure quadrupole anisotropy.

As mentioned in section~\ref{ssec:anisotropy measures}, the quadrupole distribution that will be considered in this paper 
is of the form $1-B\cos^2\theta$, aligned with a particular quadrupole axis.
The quadrupolar distribution is a linear combination of the monopole term $Y_{00}$ and $Y_{20}$ oriented 
along the quadrupole axis.  The distribution has two minima at opposite ends of the quadrupole axis and a 
maximum in the plane perpendicular to this axis.
The quadrupolar data and the reconstruction of the quadrupole axis are shown in the lower panel of Fig.~\ref{fig:skymaps}.  

For the full-sky case, the method outlined by Sommers in~\cite{Sommers:2000us} is used to reconstruct 
the quadrupole amplitude and axis.
It is possible to accurately reconstruct the quadrupole moment for experiments with partial-sky exposure at 
particular latitudes, independently of their exposure function.
This is because there is very little quadrupole moment in the exposure function, as discussed in Ref.~\cite{Denton:2014hfa}.
By some chance, Auger is exactly at the optimal latitude in the southern hemisphere, 
and TA is very close to the optimal latitude in the northern hemisphere.
Therefore we use Sommers's technique for quadrupole reconstruction of both full-sky EUSO and partial-sky Auger.

\subsection{Distinguishing Between Dipoles and Quadrupoles}
\label{ssec:distinguishing}
\begin{figure*}
\centering
\includegraphics[width=0.497\textwidth]{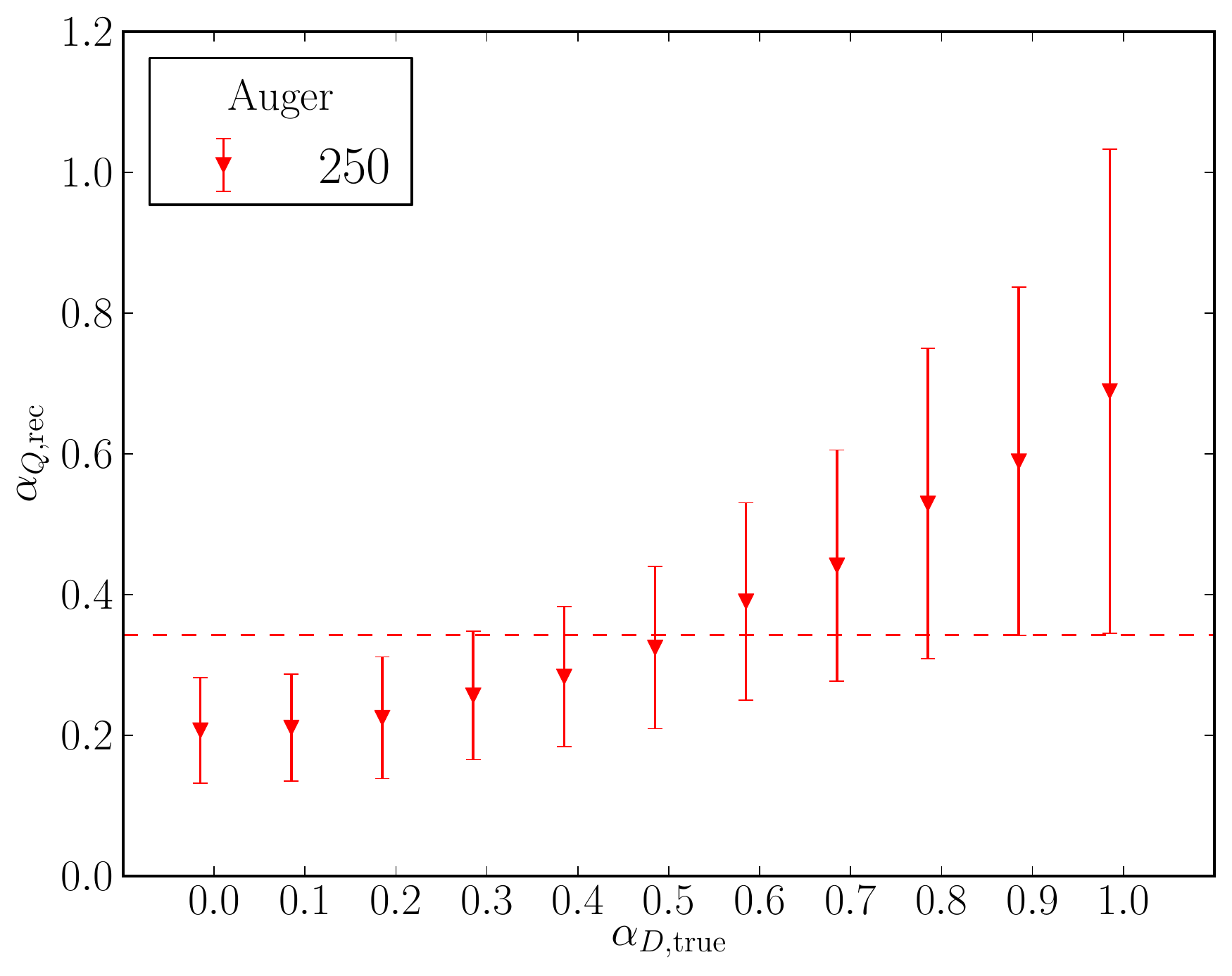}
\includegraphics[width=0.497\textwidth]{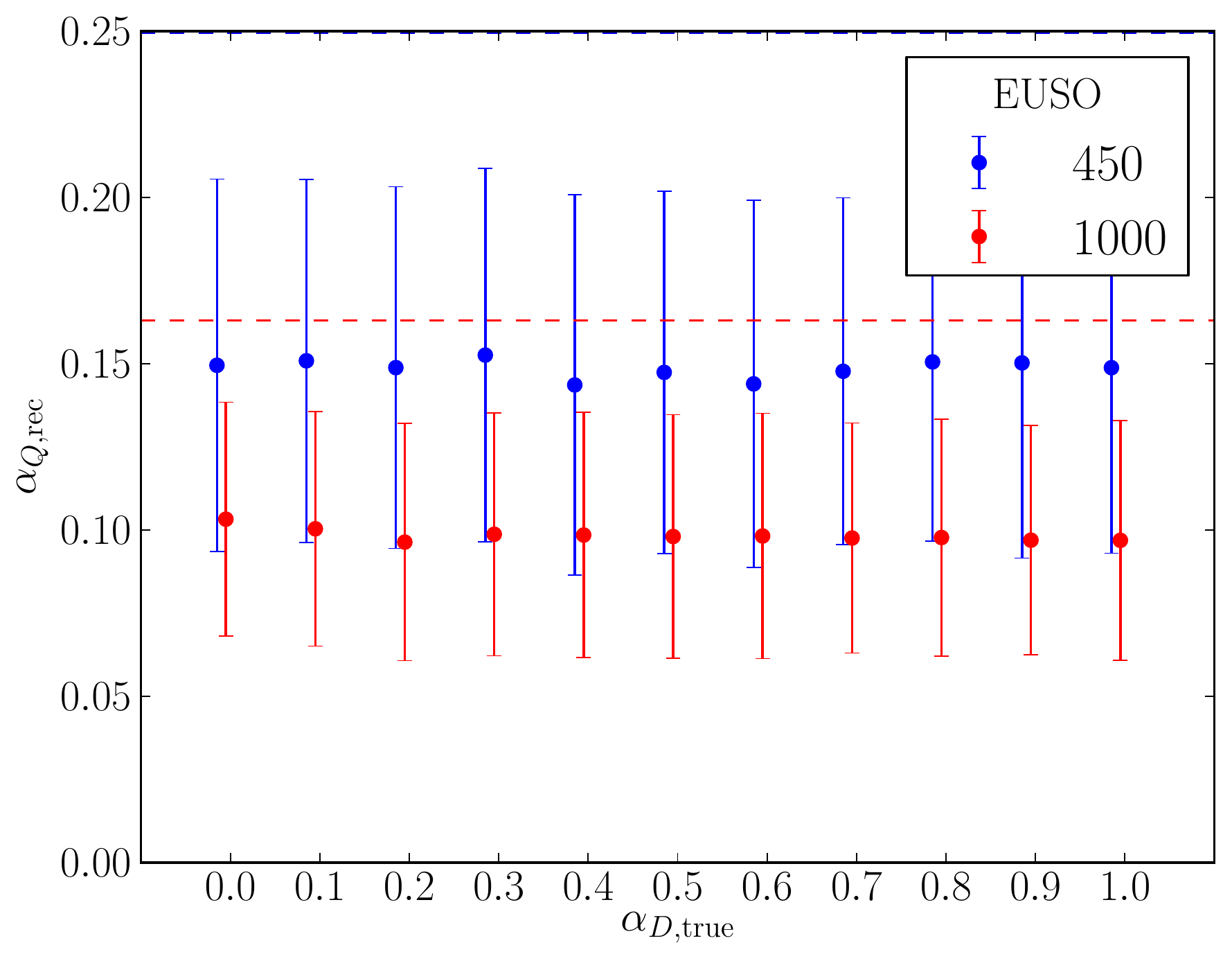}
\includegraphics[width=0.497\textwidth]{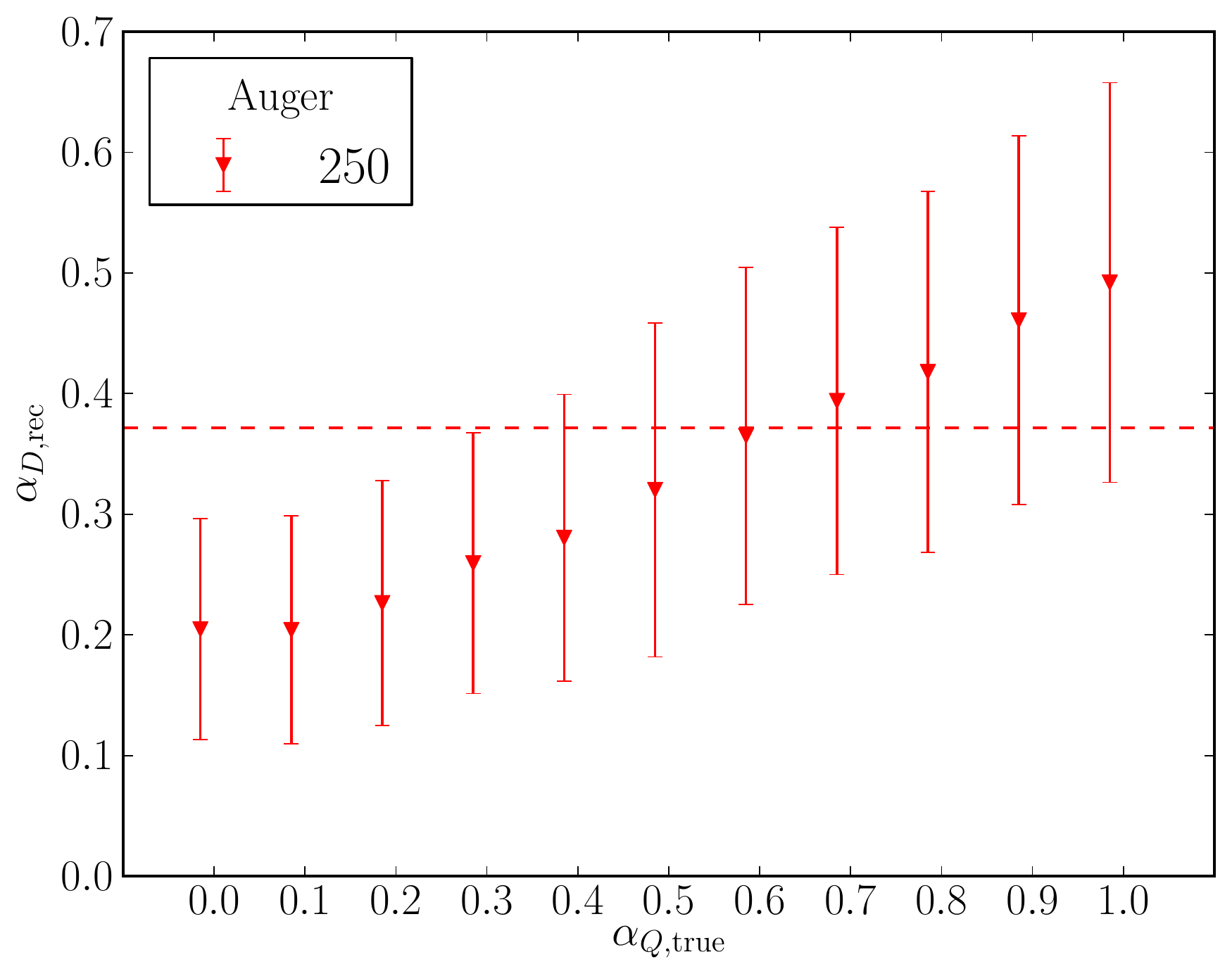}
\includegraphics[width=0.497\textwidth]{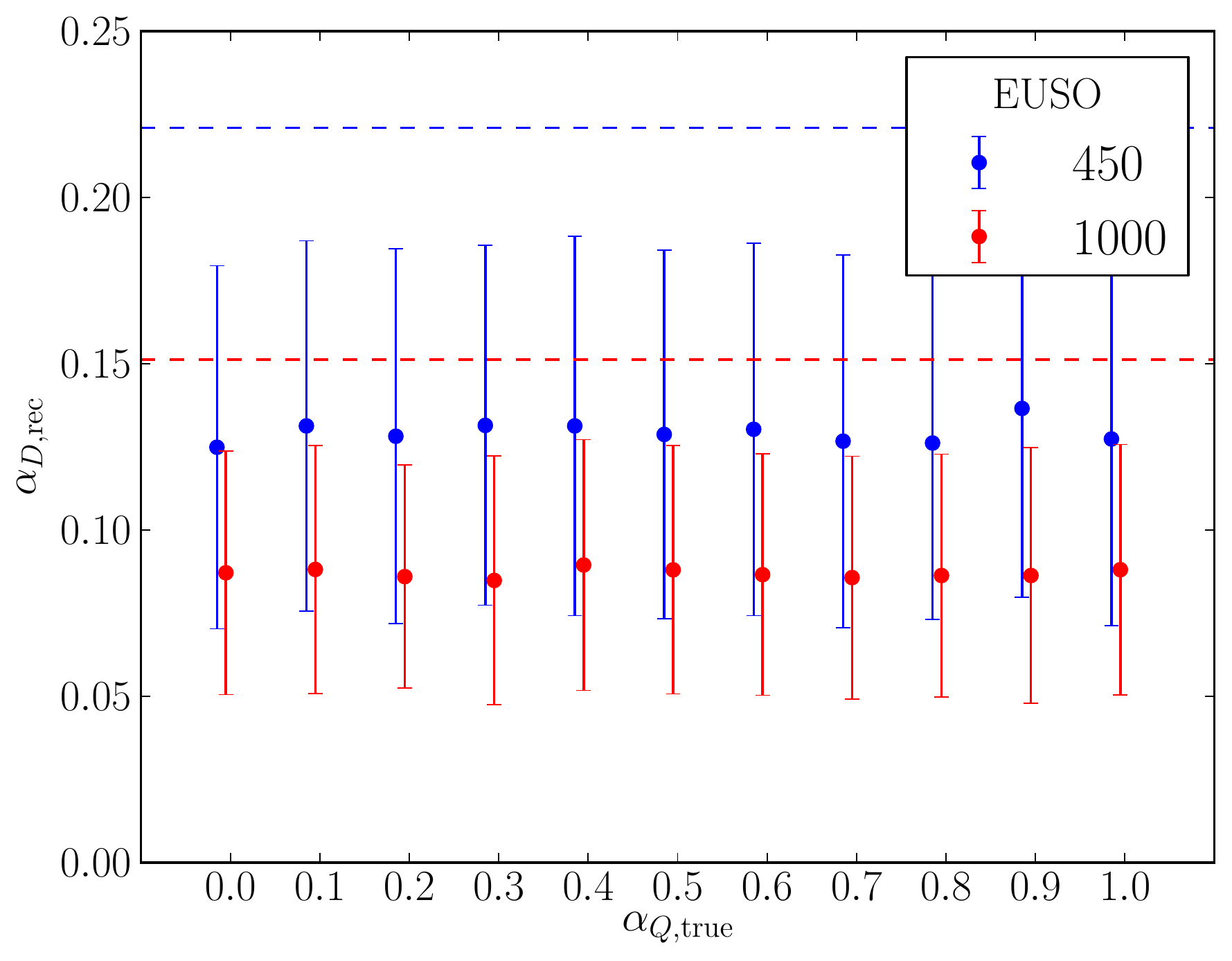}
\caption{These panels show the results of attempting to reconstruct a dipole (quadrupole) 
when there is actually a quadrupole (dipole).
The top two panels show the effect of attempting to infer a quadrupole moment from a pure dipole state 
of varying magnitudes while the bottom two panels show the effect of attempting to infer the dipole moment 
from a pure quadrupole state of varying magnitudes.
The left two panels assume Auger's partial coverage and $250$ cosmic rays, 
while the right panels assume uniform exposure and the estimated number of events 
for EUSO (450 minimally, and 1000 maximally).
The mean values and the one standard deviation error-bars are derived from 500 samplings.
Note that the left most data point in each plot ($\alpha_{(D,Q),\rm{true}}=0$) corresponds to the isotropic case,
for which the dashed lines are the 95\% upper limit.
Finally, note that the vertical scales vary significantly between the partial-sky low statistics and full-sky larger statistics
figures.
}
\label{fig:distinguishing} 
\end{figure*}

One topic of concern is determining at what significance an injected dipole (quadrupole) distribution 
can be distinguished from a quadrupole (dipole), and from isotropy.
Generally, the level of significance will depend on the number of observed cosmic-ray events, 
the strength of the anisotropy, etc. 
Fig.~\ref{fig:distinguishing} shows what happens when Auger or EUSO attempt to reconstruct a pure dipole or a
pure quadrupole when the signal is actually the opposite.
The mean values and one standard deviation error-bars are derived from 500 repetitions of the given number of cosmic-ray events, 
where the dipole or quadrupole axis direction is randomly distributed on the sphere.
The dashed lines in each plot are the 95\% upper limit for an isotropic distribution (i.e., $\alpha_{\rm true}=0$).
We see that as the actual anisotropy strength increases, quite a significant region of the parameter space would show an
anisotropy in the absent multipole at the 95\% confidence level when reconstructed by Auger.
We also see that the relative size of the error bars reflects the statistical advantage of space-based observatories, 
while the central values of the data points, falsely rising with $\alpha_{\rm true}$ for Auger but constant for EUSO,
reveals the systematic difference of partial-sky coverage versus full-sky coverage.

This entire discussion is easily understood in the context of the ``interference'' of spherical harmonics which have been effectively
truncated on the part of the sky where the exposure vanishes.
The various truncated harmonics interfere heavily, a fact that is built into the $K$-matrix method 
(and into any method that attempts to reconstruct spherical harmonics based on only partial-sky exposure).
Even though the true exposure of EUSO won't be exactly uniform, the fact that it sees the entire sky with nearly comparable coverage means that the individual spherical harmonics are non-interfering, and so can be treated independently.

\section{Results}
\label{sec:results}
In this section we tally our results.
The standard procedure involves simulating a number of cosmic rays with a given dipolar or quadrupolar anisotropy shape and amplitude
($\alpha_{\rm true}$) aligned in a random direction.
We then reconstruct the amplitude ($\alpha_{\rm rec}$) and direction (here we assume knowledge of the kind of anisotropy -- dipole or
quadrupole -- unlike in section~\ref{ssec:distinguishing}) and compare to the true values.
This process is repeated 500 times and the shown uncertainties are one standard deviation over the 500 repetitions.

\subsection{Dipole results}
\label{ssec:dipole results}
\begin{figure*}[t]
\centering
\includegraphics[width=0.497\textwidth]{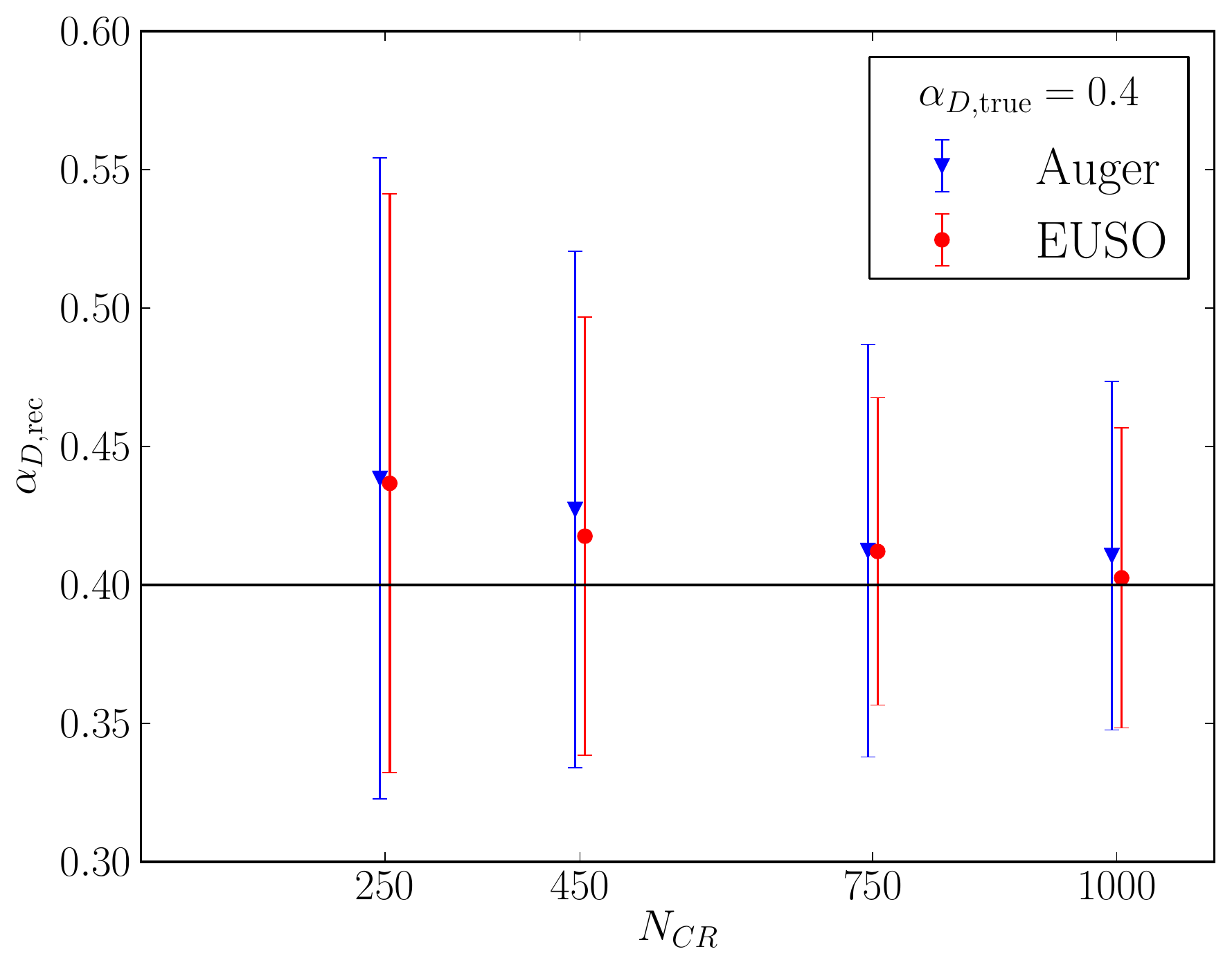} 
\includegraphics[width=0.497\textwidth]{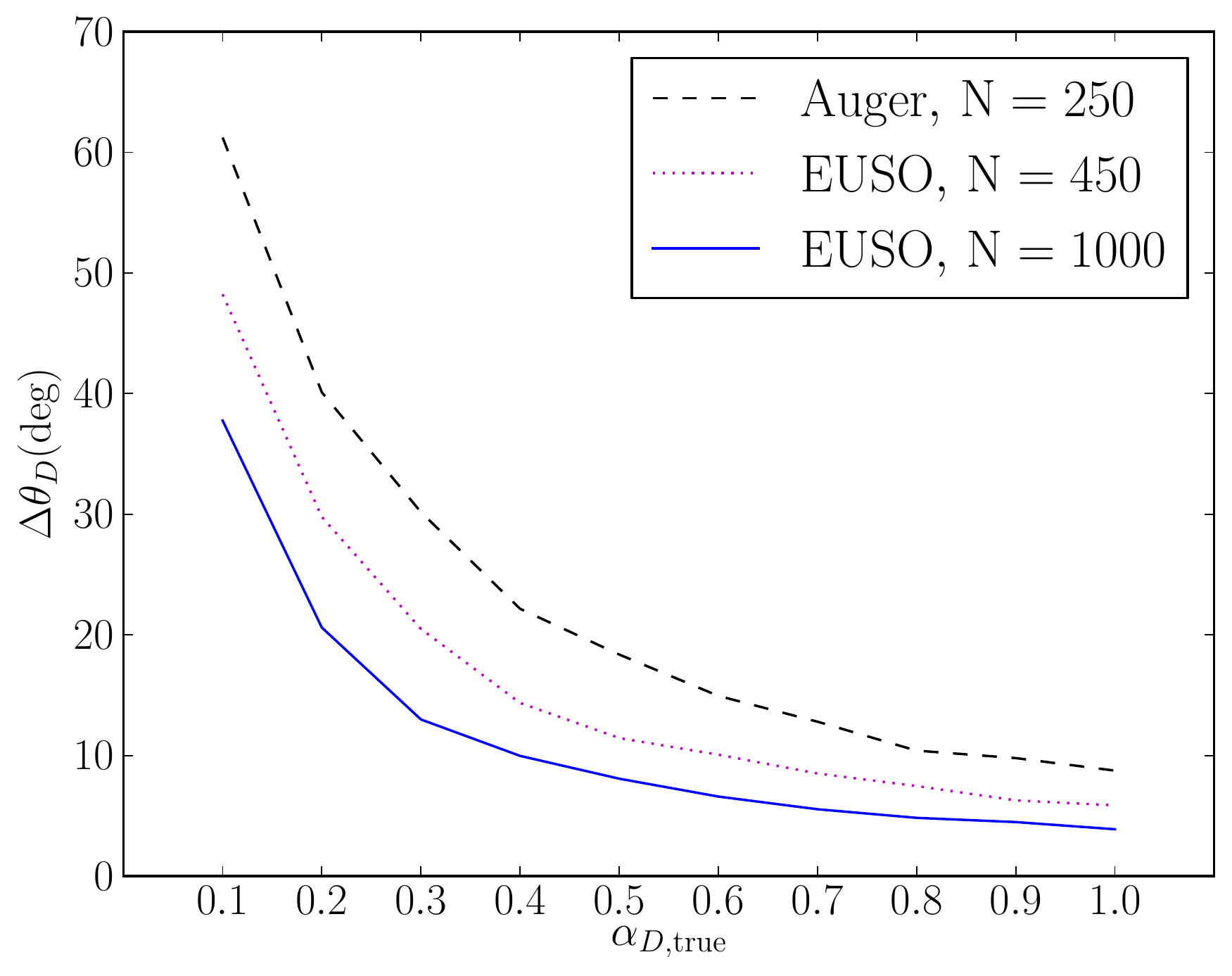}
\includegraphics[width=0.497\textwidth]{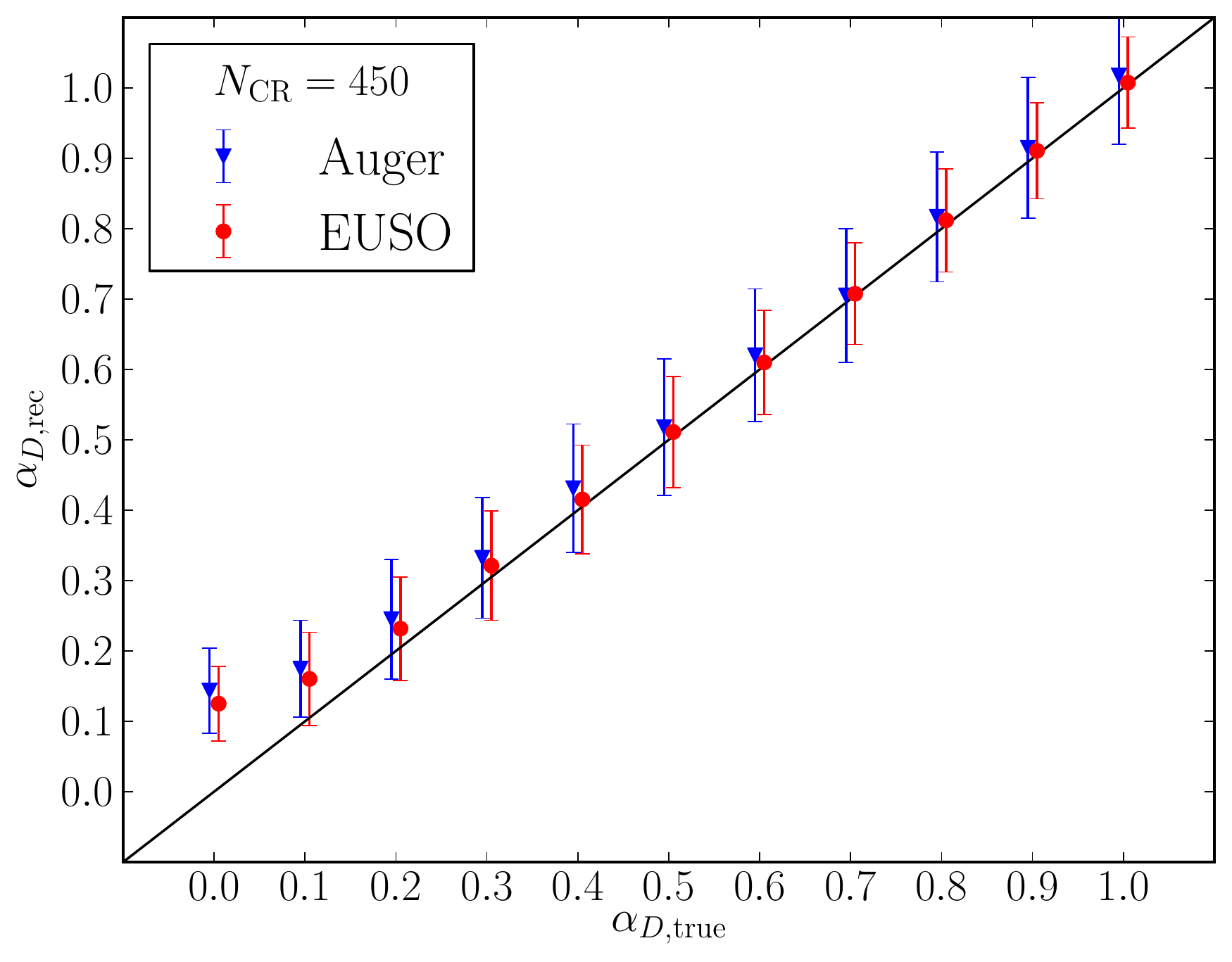} 
\includegraphics[width=0.497\textwidth]{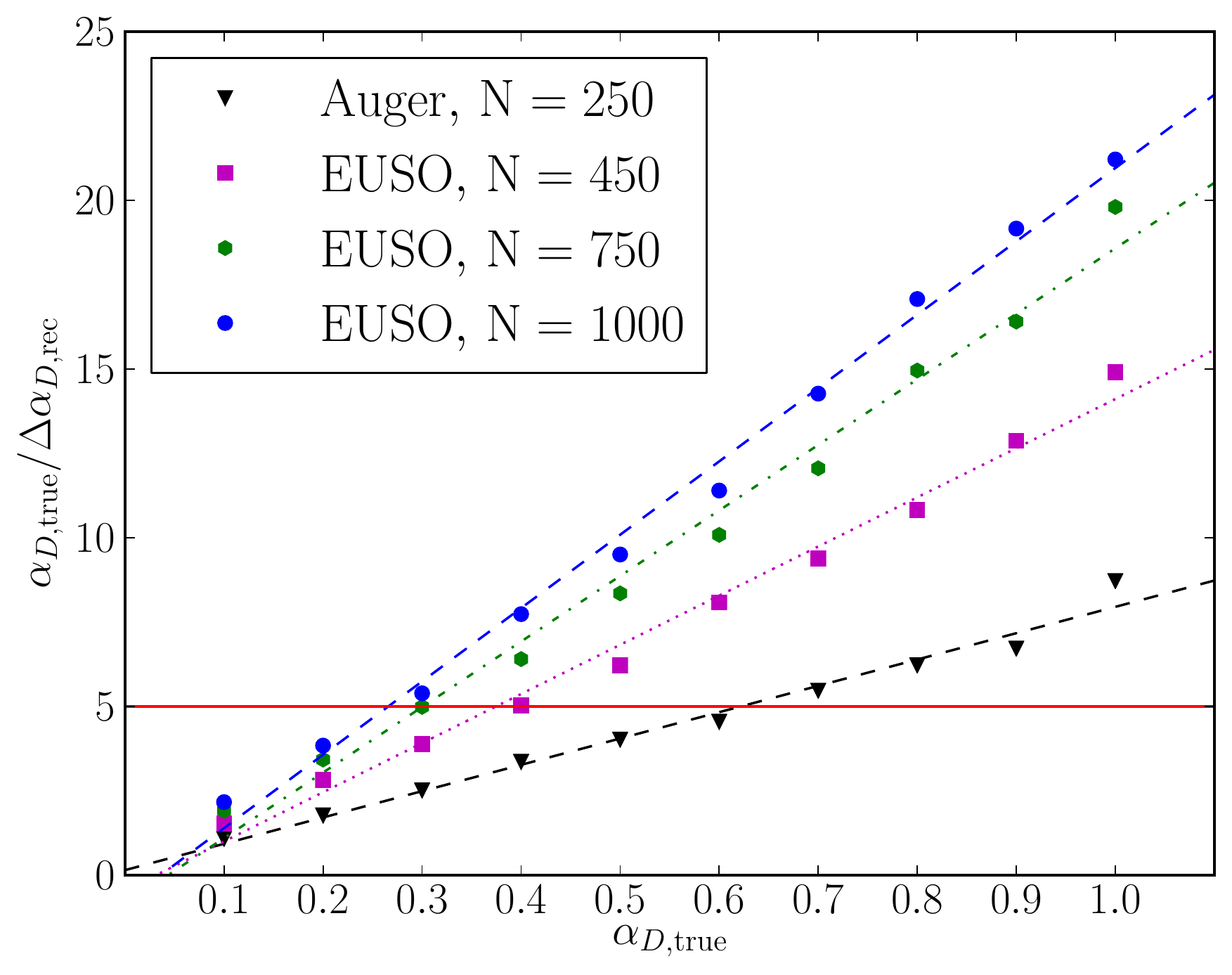}
\caption{Reconstruction of the dipole amplitude and direction across various parameters.
Each data point is the mean value (and one standard deviation error-bar as applicable) determined from 500 independent simulations.
The dipole amplitude and direction for Auger's partial coverage were reconstructed with the $K$-matrix approach.
The ordinate on the fourth panel, $\frac{\alpha_{\rm true}}{\Delta\alpha_{\rm rec}}$, 
labels the number of standard deviations above $\alpha_D=0$.}
\label{fig:dipole reconstruct}
\end{figure*}

In Fig.~\ref{fig:dipole reconstruct} we compare the capabilities of design EUSO and Auger to reconstruct a dipole anisotropy.
In this comparison, both advantages of EUSO, namely the increased FOV and the $4\pi$~sky coverage, are evident.
The first panel shows how changing only the exposure function between Auger and EUSO 
affects the value of the reconstructed dipole amplitude.
For the same number of cosmic-ray events, the EUSO reconstruction is  a bit closer to the expected value and has a smaller variation
than does the Auger reconstruction.
In the next panel we show the angular separation between the actual dipole direction and 
the reconstructed direction for Auger after a
maximal amount of Auger data of $\sim250$ cosmic-ray events, 
compared to EUSO's minimal and maximal data sizes: 
450 and 1000 cosmic-ray events, respectively.
Even for a pure dipole, Auger will only reach $10^\circ$ accuracy in dipole direction for a maximum strength dipole, 
$\alpha_D=1$, while EUSO does much better.
In the third panel we compare both experiments at the same number of cosmic rays across a range of dipole strengths.
Even if we assume that Auger will see significantly more cosmic rays than it is expected to, it still has a larger error in its
ability to reconstruct a dipole of any amplitude than EUSO.
The low dipole magnitudes will always lead to a small erroneously reconstructed dipole due to random-walking away from zero.
Finally, in the fourth panel we show the discovery power of each experiment to distinguish a dipole amplitude from isotropy.
We see that Auger with 250 events would claim a discovery at five standard deviations above isotropy only if the dipole strength is
$0.62$ or greater -- a situation that is unlikely given Auger's anisotropy results to date~\cite{Deligny:2014fxa}.
EUSO could claim the same significance if the dipole amplitude is $0.37,0.30,0.27$, or greater, for 450, 750, or 1000 events,
respectively.
The EUSO significance should be enough to probe at high significance the weak signal currently reported by Auger. 

\subsection{Quadrupole results}
\label{ssec:quadrupole results}
\begin{figure*}[t]
\centering
\includegraphics[width=0.497\textwidth]{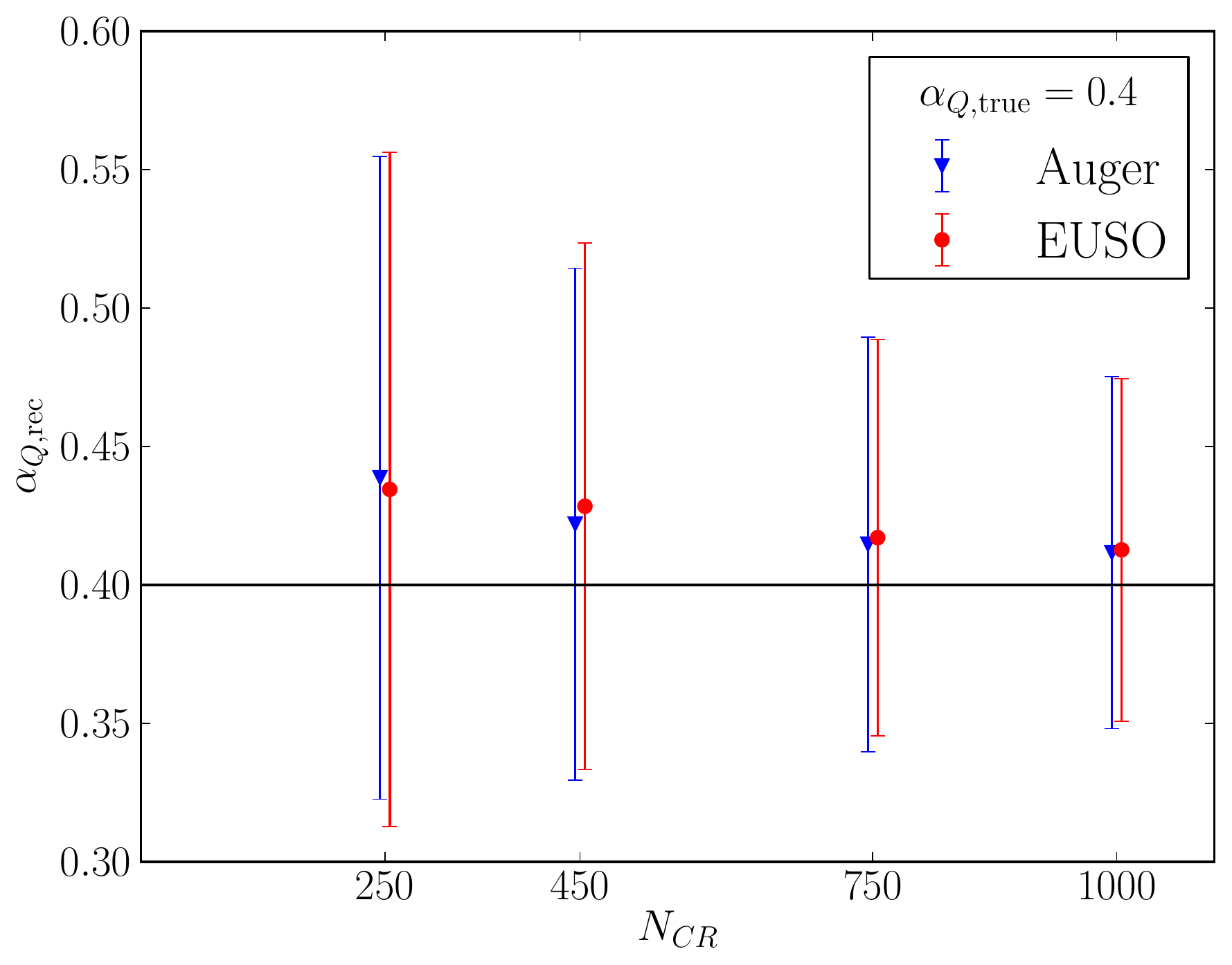}   
\includegraphics[width=0.497\textwidth]{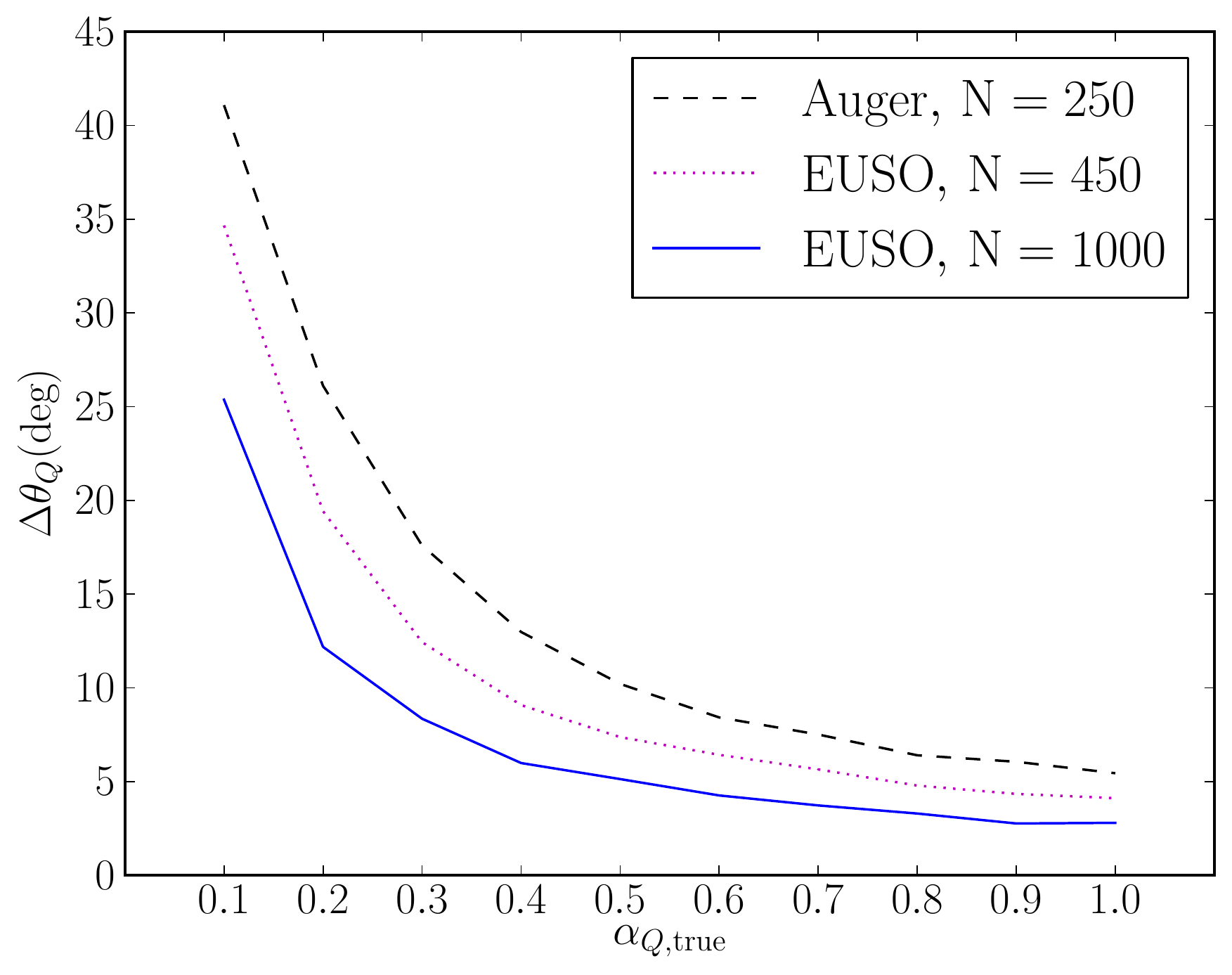}
\includegraphics[width=0.497\textwidth]{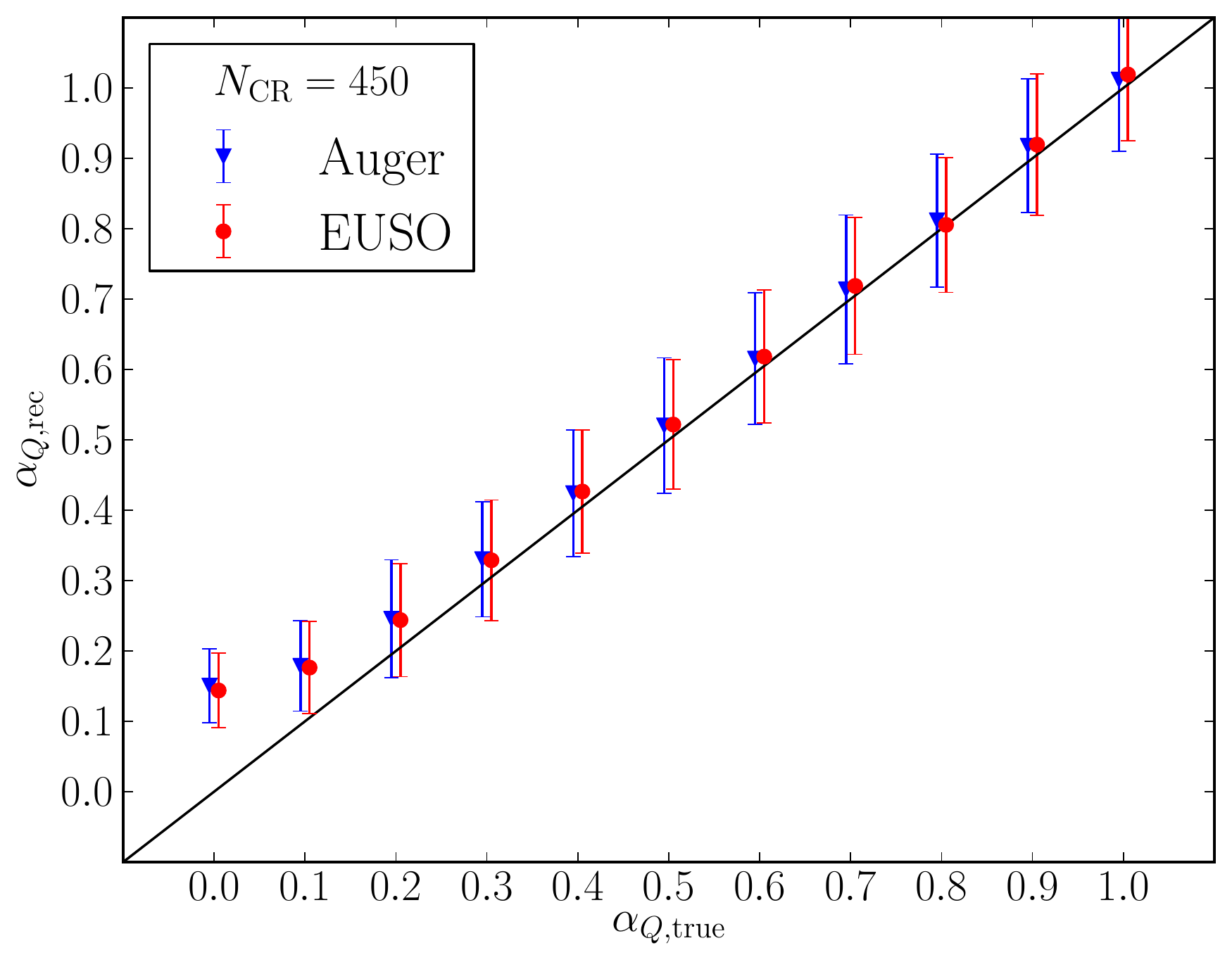} 		
\includegraphics[width=0.497\textwidth]{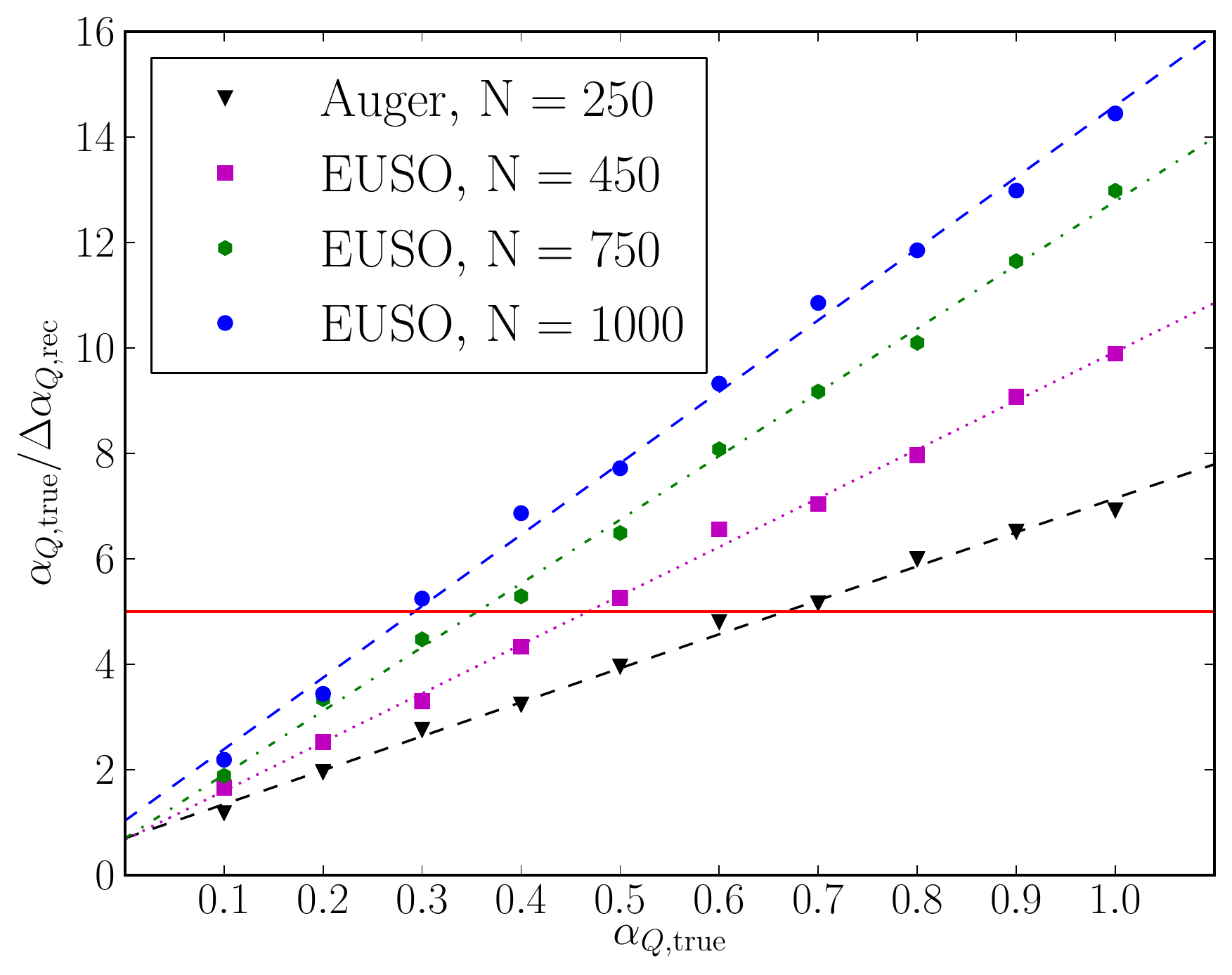}
\caption{Reconstruction of the quadrupole amplitude and direction across various parameters.
Each data point is the mean value (and one standard deviation error-bar as applicable) determined from 500 independent simulations.
The quadrupole amplitude and direction for Auger's partial coverage were reconstructed with 
the same Sommers-approach as for the full-sky (EUSO) case.
The ordinate on the fourth panel, $\frac{\alpha_{\rm true}}{\Delta\alpha_{\rm rec}}$, 
labels the number of standard deviations above $\alpha_Q=0$.}
\label{fig:quad reconstruct}
\end{figure*}

In Fig.~\ref{fig:quad reconstruct} we again compare Auger and design EUSO in the context of quadrupole anisotropies.
The same panels are plotted here as in Fig.~\ref{fig:dipole reconstruct} except with an initial quadrupole 
rather than dipole anisotropy, and a quadrupole reconstructed.
We note that while the increased number of events that EUSO will detect will certainly lead to a better resolution of the quadrupole
amplitude (as shown in the first and fourth panels) and direction (as shown in the second panel), we see that the full-sky coverage
does not provide any benefit in this case (as deduced from the first and third panels).
This result confirms the claims made in \S~\ref{ssec:reconstructing quadrupole} and in Ref.~\cite{Denton:2014hfa}.

Even though EUSO gains no benefit from its full-sky exposure for the determination of a quadrupole 
anisotropy, EUSO's increased
statistics will still lead to a detection sooner than Auger.
Auger with 250 events would only be expected to claim a quadrupole discovery at five standard deviations above isotropy
if the quadrupole strength is $0.67$ or greater.
EUSO could claim the same significance if the quadrupole amplitude is $0.47,0.36,0.29$, or greater, for 450, 750, or 1000 events,
respectively.

\section{Conclusion}
\label{sec:conclusion}
Many well motivated models predict, in the simplest limit, a dipolar or quadrupolar anisotropy in the EECR flux.
The importance of the two lowest non-trivial orders ($\ell=1,2$) can be seen from the 2MRS distribution of the 5310 
nearest galaxies that was demonstrated in Fig.~\ref{fig:Clgals}.
Due to the lack of any conclusive anisotropy from the partial-sky ground-based experiments, 
we explored the possible benefits that a full-sky space-based experiment, such as proposed EUSO, has over a ground-based experiment
for detecting dipolar or quadrupolar anisotropies.
In particular, we see that in addition to the increased statistics that proposed EUSO brings over any ground based experiment, 
proposed EUSO significantly outperforms Auger when reconstructing a dipole.
Moreover, for inferences of both the dipole and the quadrupole, partial-sky experiments fail to differentiate between the two due to
the mixing of the spherical harmonics when truncated by the exposure function.
This situation is not present with all-sky observation, where the exposure function is nearly uniform and nonzero everywhere.

\section{Acknowledgements}
\label{sec:ack}
We have benefited from several discussions with L.A.~Anchordoqui.
This work is supported in part by a Vanderbilt Discovery Grant.

\bibliography{UHECR}

\end{document}